\documentclass[aps,prl,superscriptaddress,showpacs,nofootinbib,reprint,floatfix]{revtex4-1}
\usepackage{mciteplus}
\usepackage{latexsym}
\usepackage{amsthm}
\usepackage{amsmath}
\usepackage{amssymb}
\usepackage{hepunits}
\usepackage{hyperref}
\usepackage{bbm}
\usepackage{bm}
\usepackage{xfrac}
\usepackage{color}
\usepackage{colordvi}
\usepackage{comment}
\usepackage{dcolumn}
\usepackage{times,latexsym,graphicx,wrapfig,url}
\usepackage{epsfig,lineno,bm}
\usepackage{subfigure}
\usepackage{slashed}

\def\apr{A^\prime}

\newcommand{\be}{\begin{eqnarray}}
\newcommand{\ee}{\end{eqnarray}}
\newcommand{\bea}{\begin{eqnarray}}
\newcommand{\eea}{\end{eqnarray}}

\newcommand{\bef}{\begin{figure}[htbp]\begin{center}}
\newcommand{\eef}{\end{center}\end{figure}}
\newcommand\Perimeter{Perimeter Institute for Theoretical Physics, Waterloo, ON N2L 2Y5, Canada}

\def\lsim{\mathrel{\rlap{\lower4pt\hbox{\hskip1pt$\sim$}}
    \raise1pt\hbox{$<$}}}
\def\gsim{\mathrel{\rlap{\lower4pt\hbox{\hskip1pt$\sim$}}
    \raise1pt\hbox{$>$}}} 

\begin{document}

\title{Accelerating the Discovery of Light Dark Matter}
\author{Eder~Izaguirre}          \thanks{eizaguirre@perimeterinstitute.ca}  \affiliation{\Perimeter} 
\author{Gordan~Krnjaic}          \thanks{gkrnjaic@perimeterinstitute.ca}  \affiliation{\Perimeter} 
\author{Philip~Schuster}           \thanks{pschuster@perimeterinstitute.ca}  \affiliation{\Perimeter} 
\author{Natalia~Toro}          \thanks{ntoro@perimeterinstitute.ca}  \affiliation{\Perimeter} 

\date{\today}
\begin{abstract}
We analyze the present status of sub-GeV thermal dark matter annihilating through Standard Model mixing and identify a small set of future experiments that can decisively test these scenarios.  
%
\end{abstract}


\maketitle

\section{Introduction}


That dark matter (DM) is a thermal relic from the hot early Universe is an 
inspiring possibility that motivates non-gravitational interactions between dark and ordinary matter.  
The canonical example involves a heavy particle interacting through the weak force (WIMPs). 
This scenario has motivated searches for DM scattering in underground 
detectors, for DM annihilation in the cosmos, and for DM production in high-energy colliders. These efforts achieve broad and powerful sensitivity to DM with mass between a few GeV and the TeV scale.

A thermal origin is equally compelling --- and, in simple models, predictive --- even if DM is not a WIMP.
 DM with any mass from an MeV to tens of TeV can achieve the correct relic abundance by annihilating directly into Standard Model (SM) matter.  However, the lower half of this  mass range cannot be fully explored using existing strategies -- an unfortunate situation that jeopardizes the legacy of the DM search effort. In particular, DM-nuclear and DM-electron scattering searches lose sensitivity precipitously for DM lighter than a few GeV or DM that scatters inelastically;
limits on DM annihilation at low temperatures (most notably from the CMB) are irrelevant to many scenarios; and missing energy searches at high-energy 
colliders are blind to the interactions responsible for the DM abundance.

In this paper, we sharply define the challenge of testing sub-GeV thermal DM.   Below a few GeV, avoiding DM overproduction requires that a comparably light particle mediate DM annihilation  \cite{Lee:1977ua,Boehm:2002yz,Boehm:2003hm,Pospelov:2007mp, Pospelov:2008zw,ArkaniHamed:2008qn,secluded-wimps}.  We focus on the case where DM annihilates through an $s$-channel mediator directly into SM states\footnote{Sufficiently heavy DM can also annihilate into two mediators --- a far less predictive scenario.}. 
We classify annihilation mechanisms arising from dark-sector mixing with SM fields, define minimal models representative of each mechanism, and compute all relevant constraints on each model.  We introduce a framework to compare all such constraints to one another and to the milestone provided by thermal DM freeze-out.
Finally, we identify a small set of ``flagship'' experiments with complementary sensitivity --- using direct detection, $B$-factory mono-photon, and fixed-target missing momentum strategies --- that together can decisively test light DM annihilating through SM mixing.

\section{GeV-Scale Dark Matter}
Viable light thermal DM scenarios can be classified by the spins and masses of the DM and mediator, by whether  the thermal abundance or a primordial asymmetry dominates the DM density, and by the mediator's interactions with both DM and SM matter.  SM symmetries substantially restrict the latter interactions: vector mediators can mix with the photon or weakly gauge a SM global symmetry, while scalars can mix with the Higgs (or have axion-like couplings in extensions of the SM).  
Rare $B$-meson decays largely exclude the scalar mediator scenarios for sub-GeV thermal DM (see below), so we focus here on the possibilities  for DM coupled through a vector mediator. 

 One concrete example is a scalar QED model of DM, 
where the ``dark photon'' $A'$ is massive with coupling $g_D \equiv \sqrt{4\pi \alpha_D}$ to scalar DM 
currents ${\cal J}_D^\mu=i\varphi^*\partial^{\mu}\varphi+c.c.$. The DM 
scalar $\varphi$ and dark photon $A'$ have masses $m_{\varphi}$ and $m_{A'}$ respectively. 
The leading SM coupling to this dark sector allowed by symmetries is 
photon-$A'$ kinetic mixing, ${\cal L}_{ mix}=\epsilon {F^\prime}^{\mu\nu}F_{\mu\nu}$ where ${F^\prime}$
and ${F}$ are the $A'$ and photon field strengths, respectively \cite{Okun:1982xi,Holdom:1985ag}. 
After diagonalizing away the kinetic mixing term, the low-energy Lagrangian 
that describes dark-visible interactions is
\be
{\cal L}_{int}= A^\prime_\mu ( \epsilon e {\cal J}^{\mu}_{EM} + g_D {\cal J}_D^\mu), 
\label{eq:vectorportal}
\ee 
where ${\cal J}^\mu_{EM}$ is the SM electromagnetic current. 


\begin{figure*}[t] 
\vspace{0.1cm}
\hspace{-0.4cm}

\includegraphics[width=5.5cm]{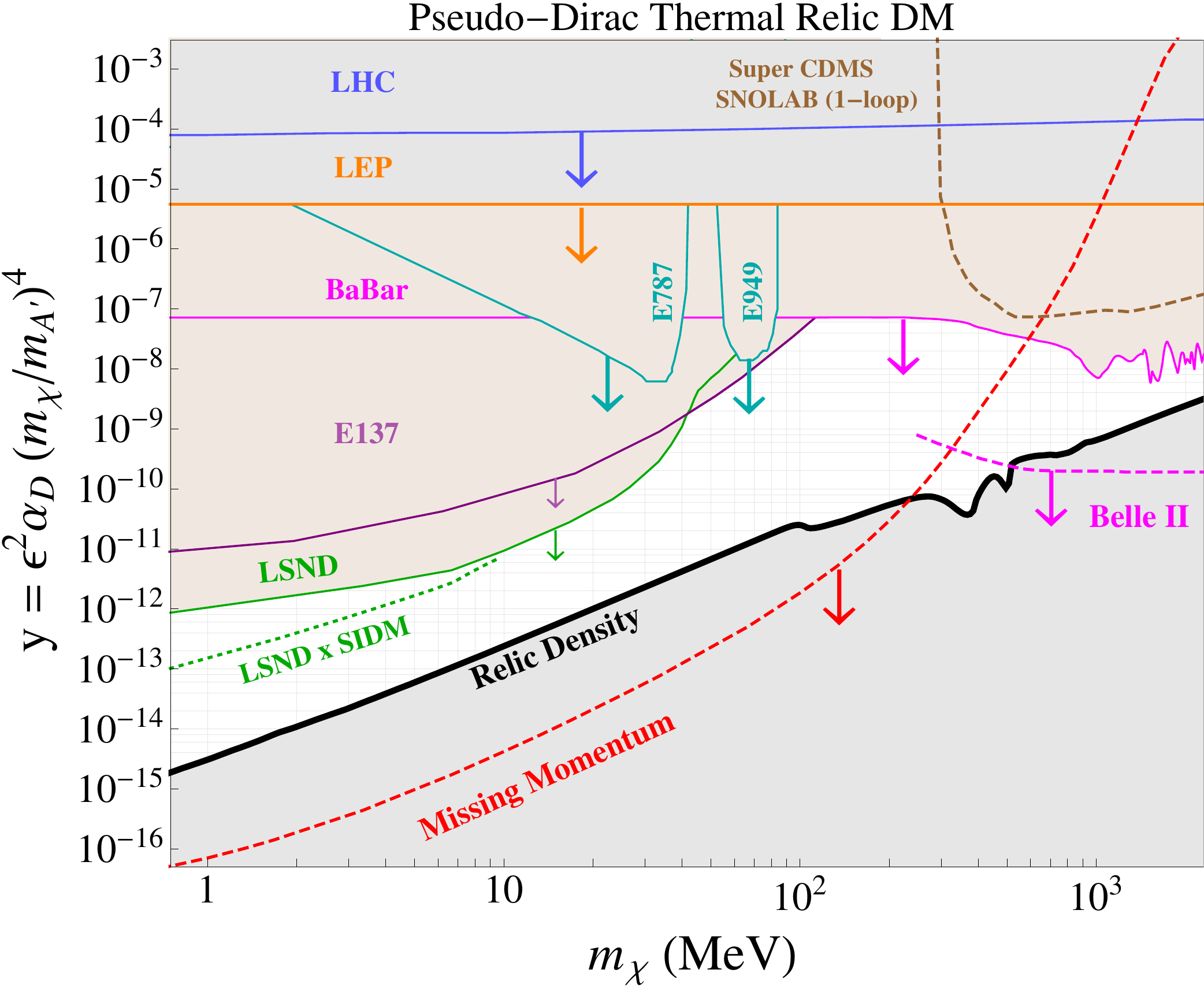}  ~~
\includegraphics[width=5.5cm]{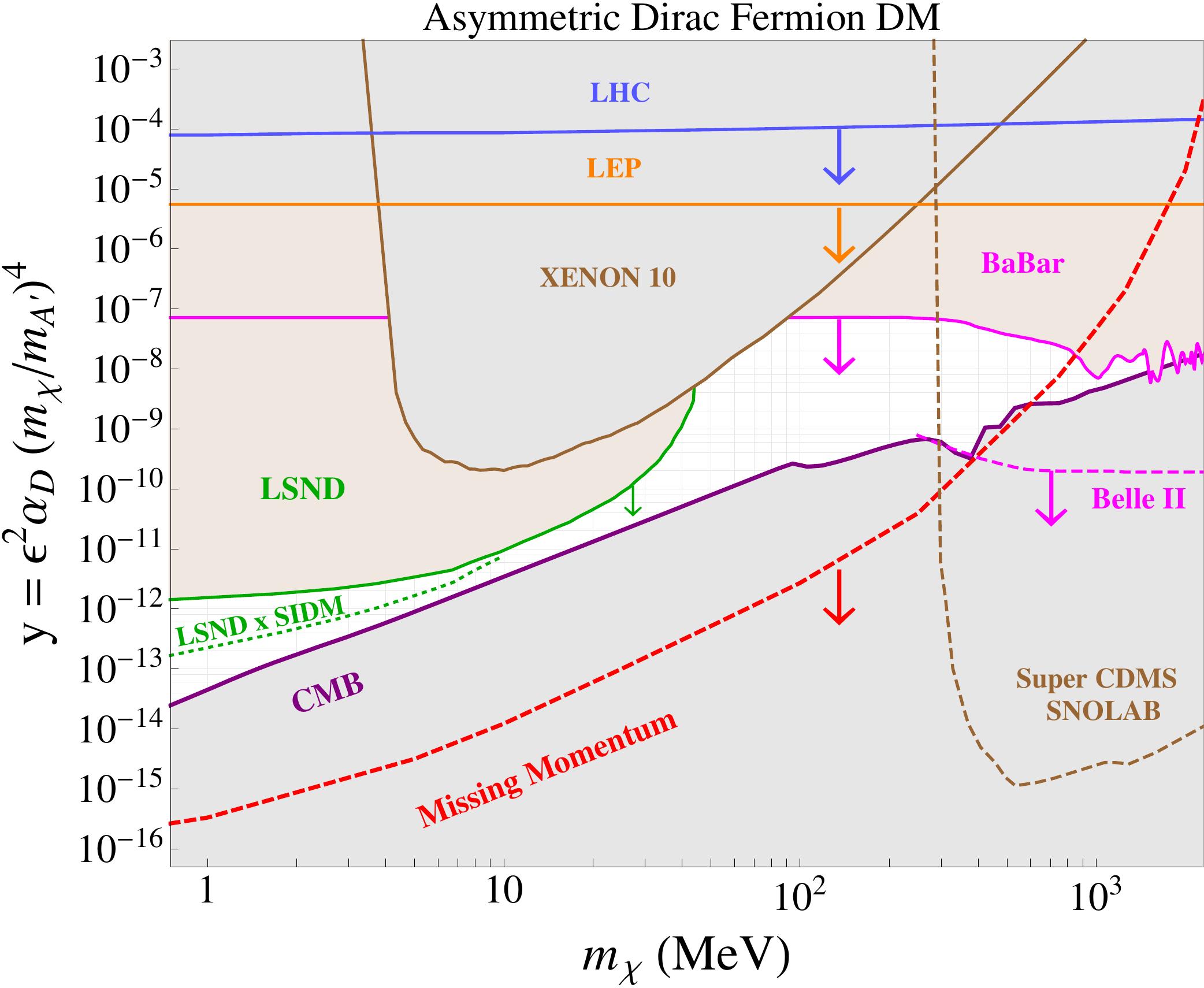}  ~~~
\includegraphics[width=5.5cm]{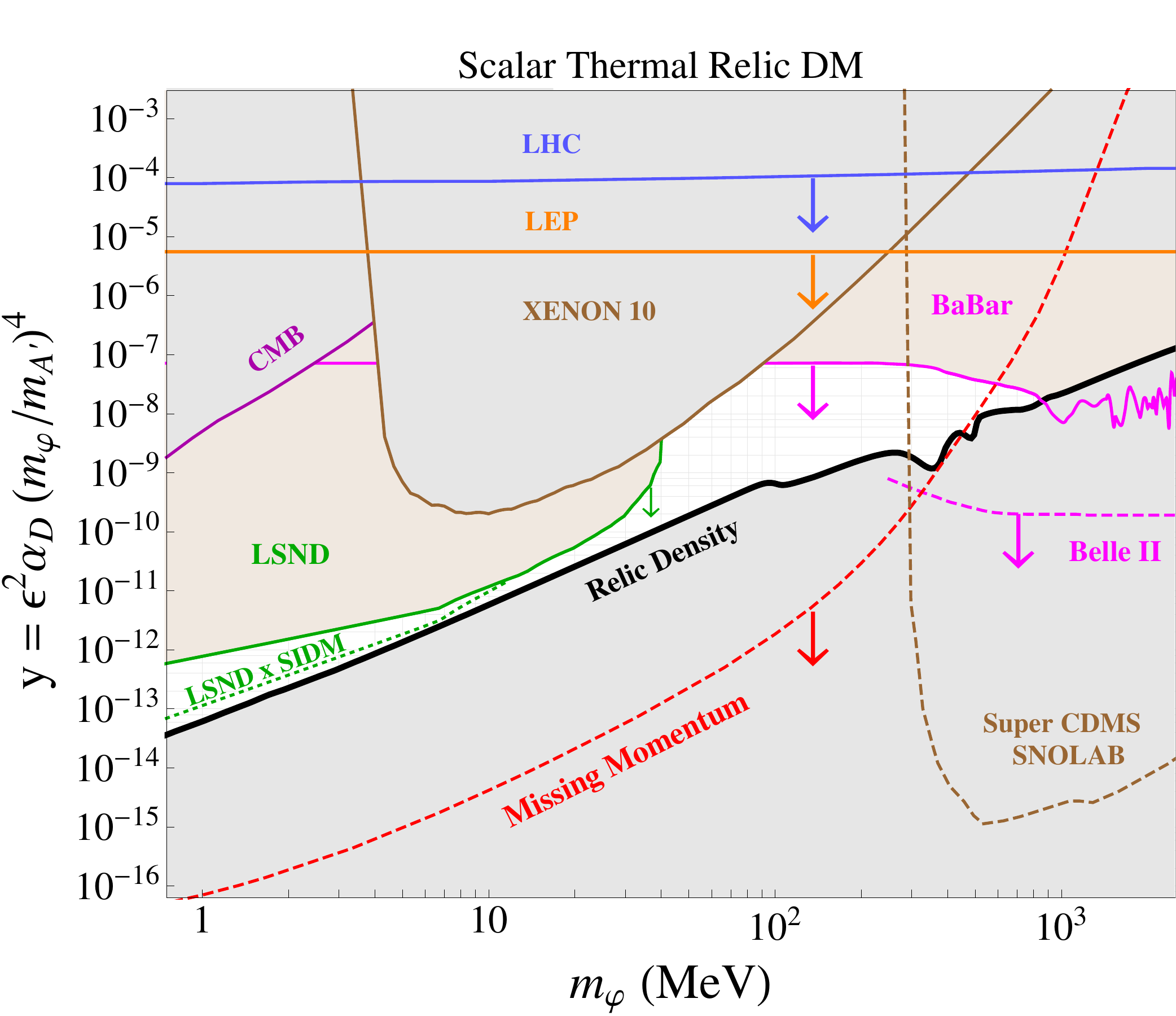}  ~~~

  \caption{
  Constraints and projections for representative vector-portal DM scenarios. For definiteness, we evaluate all constraints for $m_{\rm DM} /m_{\apr}  = 1/3$ and (except for the LSND$\times$SIDM bound -- see below), $\alpha_D = 0.5$, near the perturbativity limit.
   The relic density, CMB, and direct detection contours scale roughly as $\epsilon^2 \alpha_D (m_{\rm DM}/m_{\apr})^4$ (plotted on the $y$-axis), and so are insensitive to separate factors in the above.  For other constraints, this choice is conservative, in that smaller choices of $\alpha_D$ and/or $m_{\rm DM} /m_{\apr}$ would shift the shaded regions downward (see text); arrows denote the shift in sensitivity for $\alpha_D \rightarrow 0.05$.  
We illustrate these constraints for {\bf (left)} pseudo-Dirac/inelastic fermion thermal-relic DM, with splitting $\delta \gtrsim100 \, \keV$, {\bf(center)} asymmetric Dirac fermion DM, and {\bf(right)} scalar elastic-scattering thermal relic DM. Dirac fermion thermal-relic DM is fully excluded by the CMB constraint and inelastic or asymmetric scalar DM is quite similar to the right figure, but with CMB and direct detection constraints weakened.
CMB, self-interaction (SIDM), and direct detection constraints all depend on the $\chi(\varphi)$ abundance, and are computed assuming the full DM abundance, \emph{not} the thermal abundance expected for given masses and couplings. 
In all plots,  gray shaded regions (color online) 
  represent  traditional DM constraints (e.g. direct detection), while non-traditional accelerator probes are shaded beige. We note that pseudo-Dirac limits are modified (and new dedicated searches are possible \cite{OurFuturePaper}) if $\delta$ is large enough that $\chi_+$ can decay on detector length-scales.}
   \label{fig:money}
\vspace{0cm}
\end{figure*}

For this representative case, we can now ask what parameter  ranges achieve the correct $\varphi$ relic density. 
For $m_{A'}\gsim m_{\varphi}$, the rate of annihilations $\varphi\varphi^*\rightarrow \bar f f$ determines the relic density.  Neglecting $m_f / m_{\varphi}$ corrections, the tree-level annihilation cross section at relative velocity $v_{\rm rel} \ll c$ is  
\bea
\sigma v_{\rm rel} &=&    \frac{8\pi}{3} \frac{ \epsilon^2 \alpha \alpha_D m_\varphi^2 v_{\rm rel}^2}{(m_{A'}^2 - 4 m_\varphi^2)^2  +m_{\apr}^2\Gamma^2   }~  \label{eq:scalar-annihilation-rate-A},
\eea
where $\Gamma$ is the $A'$ width.  In the limit  $m_{\apr} \gg m_\varphi,\Gamma$, this cross-section depends on dark-sector parameters only through the DM mass $m_\varphi$ and the dimensionless combination 
\bea
y &\equiv& \epsilon^2 \alpha_D \left(\frac{m_\varphi}{m_{A'}}\right)^4,
\label{eq:scalar-annihilation-rate-B}
\eea
so matching the $\varphi$ relic abundance to the observed DM density essentially fixes $y$ as a function of $m_\varphi$ (models with larger $y$ can give rise to a subdominant component of the DM).  Of course, near the fine-tuned  region $m_{\apr} \approx 2 m_{\varphi}$, the precise milestone differs from that inferred from Eq.\,\eqref{eq:scalar-annihilation-rate-B}.

Before comparing existing data to this milestone, we comment on 
obvious and important variants of the model above. 
First, the DM may be a fermion instead of a scalar.  A Dirac fermion $\chi=(\chi_1,\chi^\dagger_2)$ (decomposed here into Weyl spinors) can couple to the $\apr$ through vector and/or axial currents.  The axial piece leads to velocity-suppressed $p$-wave annihilation with scaling similar to Eq.~\eqref{eq:scalar-annihilation-rate-A}, while the vector current ${\cal J}_{D}^\mu=\chi^\dagger_1\bar{\sigma}^{\mu}\chi_1-\chi^\dagger_2\bar{\sigma}^{\mu}\chi_2$ leads to $s$-wave annihilation, and typically dominates.  For this reason, we shall focus on the pure vector coupling.  


If the global symmetry under which $\chi_{1,2}$ have opposite charges
is broken (e.g. by a higgs field that gives mass to the $\apr$), operators such as ${\cal L}_{break} = \delta \chi_{1}\chi_{1}$ yield mass eigenstates $\chi_{\pm}=1 / \sqrt{2}(\chi_1\pm\chi_2)$  split in mass by $\delta$, with off-diagonal $A'$ couplings ${\cal L}_{int}= A^\prime_\mu\chi^\dagger_+\bar{\sigma}^{\mu}\chi_-$. 
This exemplifies the {\it inelastic} or {\it pseudo-dirac} scenario \cite{TuckerSmith:2001hy}.  
Analogously inelastic interactions can also arise in the scalar case. 

Finally, for either scalar or fermionic DM,  its total abundance may be set by a primordial particle/anti-particle asymmetry that dominates over the thermal relic abundance. In this case
Eq.\,\eqref{eq:scalar-annihilation-rate-A} sets  a lower bound 
on the collective interaction strength so that the symmetric component is sub-dominant. 

Each scenario  above has a counterpart where the $\apr$ couples to a global symmetry current of the SM (e.g. baryon minus lepton number), rather than via kinetic mixing.  The results that follow rely mainly on 
the $\apr$ coupling to electrons, and so apply equally well (with $O(1)$  corrections to the thermal relic  curve) to these scenarios, unless the $\apr$ gauges a symmetry under which electrons are neutral, such as $\mu -\tau$ number \cite{Altmannshofer:2014cfa,Altmannshofer:2014pba}. 

{\it \bf Scalar Mediators}
To illustrate the strong meson-decay constraints on scalar mediators, we consider one explicit model: a scalar mediator $\Phi$ that mixes with the Higgs boson and couples to a DM fermion $\chi$, with ${\cal L}_{\rm int} = \sum_i \epsilon_\Phi  y_i   \Phi  \bar f_i f_i  + y_\chi \Phi  \bar\chi \chi$, where $y_\chi$ and $\epsilon_\Phi$ are free parameters and the $y_i$ are SM Higgs Yukawa couplings, $y_i = \sqrt{2} m_i/v$ with $v=246 \, \GeV$. 
Such a $\Phi$ can mediate the partly invisible $B$-meson decays $B^+ \rightarrow  K^{+(*)} \Phi \rightarrow K^{+(*)} \chi\bar\chi$, with a rate computed (for on-shell $\Phi$) in \cite{Dolan:2014ska,Bird:2004ts}.  When  $m_\Phi > M_B - M_K \gg m_\chi$, this process (with off-shell $\Phi$) has similar kinematics to $B^{+}\rightarrow  K^{+(*)} \nu \bar\nu$, the limit on the latter \cite{Lees:2013kla} implies 
$y_\chi^2 y_t^2 \epsilon_\Phi^2 /m_{\Phi}^4 \lesssim 1.6 \times 10^{-6} \, \GeV^{-4}$.  The DM annihilation rate scales similarly, but with $y_t$ replaced by the much smaller electron and muon Yukawas.  This bound rules out thermal-relic DM for $m_\chi \lesssim  \, \GeV$.  The limits for lighter $\Phi$ and on scalar DM are even stronger, and constraints on axion-like couplings to Standard Model matter are comparable within $O(1)$ factors.   We defer a complete discussion of these scenarios for future work \cite{OurFuturePaper}.  

\section{Existing Data Confronts Light DM}
Returning to the representative scenarios with a vector mediator, we now assess how well they are constrained by current data.  
Fig. \ref{fig:money} quantifies each constraint in the plane of $y\equiv \epsilon^2 \alpha_D \left(m_\varphi / m_{A'}\right)^4$
vs. $m_\varphi$ (or similarly for a fermion $\chi$), to facilitate comparisons with the relic abundance target.
The scalings below apply for $m_{\apr} > 2m_{\varphi(\chi)}$, where the $\apr$ decays invisibly into $\varphi\,(\chi)$ pairs, but the same experiments also constrain $\varphi\,(\chi)$ production through a lighter off-shell $\apr$.  


{\it \bf CMB and BBN}
Although DM annihilations freeze out before the era of recombination, 
residual annihilations can re-ionize hydrogen and distort the high-$\ell$ CMB power spectrum \cite{Finkbeiner:2011dx, Lin:2011gj, Galli:2011rz, Madhavacheril:2013cna, Ade:2013zuv}. These data constrain the total power injected by DM annihilations \cite{Ade:2013zuv}, which scales as the DM annihilation cross-section (and hence approximately as $y$) times a correction factor calculated in \cite{Madhavacheril:2013cna}.
The resulting constraint on $y$ definitively rules out thermal-relic Dirac fermion DM (not shown in Fig. \ref{fig:money}), but not the other scenarios.  For scalar DM (Fig.~\ref{fig:money}, right), $p$-wave suppression of annihilations at late times results in a weak upper bound on $y$; the bound is likewise exponentially weakened for inelastic DM as the excited state can decay or be thermally depopulated before recombination \cite{IDMcosmology} (not shown in Fig.~\ref{fig:money} left).  For asymmetric fermion DM (Fig.~\ref{fig:money} center), CMB constraints imply a {\it minimum} value of $y$ needed to depopulate the symmetric DM component  \cite{Lin:2011gj}.  
Weaker bounds arise from the $^3$He abundance during BBN \cite{Hisano:2008ti, Hisano:2009rc,Jedamzik:2009uy, Henning:2012rm}. 

{\it \bf Direct Detection} 
Light DM scattering at halo velocities induces nuclear recoils below current experiments' energy thresholds, but electron-scattering of sub-GeV DM (whose rate scales with $y$) is constrained by XENON10 data \cite{Essig:2012yx,Essig:2011nj}. 
For pseudo-Dirac DM (Fig. \ref{fig:money}, left), tree-level scattering is inelastic and 
kinematically forbidden for $\delta \gsim \keV$; elastic scattering arises from a one-loop box diagram scaling as $y^2$.

\begin{figure}[t] 
\vspace{0.1cm}
\hspace{-0.4cm}
\includegraphics[width=8cm]{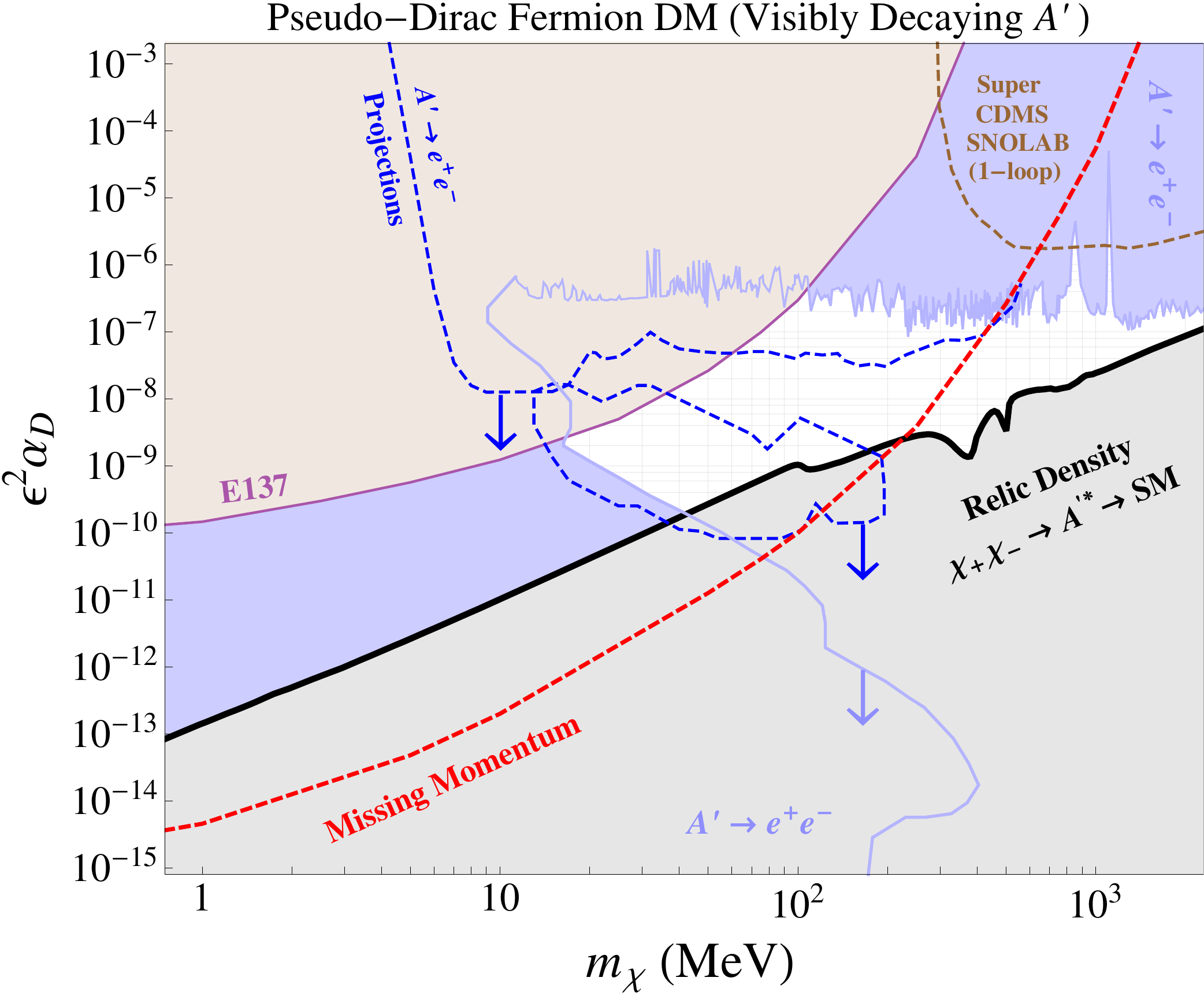}  ~~~
  \caption{Same as Fig \ref{fig:money}, but assuming $ m_{\apr}  = 1.1 m_{\chi}$, where
   the relic abundance is still achieved through 
  $s-$channel annihilation, but $\apr$ decays visibly. Dashed blue contours show projected sensitivities from \cite{Essig:2013lka} for future $\apr \to \ell^+\ell^-$ searches in the next few years.  Depending on the $\chi$ and $\apr$ masses, models below the relic density line may still achieve viable thermal abundances through other processes --- see \cite{JoshAndRafaelle}.}
   \label{fig:visible}
\vspace{0cm}
\end{figure}

{\it \bf B-Factories}
Following \cite{Izaguirre:2013uxa,Essig:2013vha}, we reinterpret the BaBar search for an (untagged) 
$\Upsilon(3S)\rightarrow \gamma+\rm{invisible}$ \cite{Aubert:2008as}
to constrain the continuum process $e^+e^-\rightarrow \gamma+A'^{(*)}\rightarrow  \gamma \chi\bar\chi$.   
For on-shell $A'$, the event rate scales as $\epsilon^2$, independent of 
$\alpha_D$, and insensitive to $m_{A'}$ and $m_\varphi$ in our kinematic regime (except for a shelf at $m_{A'}\sim 1~\GeV$ due to a peaking background).  
To compare to the relic density target in the $y\equiv \epsilon^2 \alpha_D \left(m_\varphi / m_{A'}\right)^4$ 
vs. $m_\varphi$ plane, we must fix these other parameters. To do so conservatively, we use the benchmarks 
 $m_\varphi / m_{A'}=1/3$ and $\alpha_D=1/2$ (smaller ratios and smaller $\alpha_D$ would overstate the exclusion of thermal relic DM by these results, while larger ratios qualitatively change the physics and larger $\alpha_D$ can run towards the non-perturbative regime \cite{Davoudiasl:2015hxa}).  
It can be argued that $m_\varphi\sim m_{A'}$ is most natural in any case, and the resulting constraint 
only improves linearly with decreasing $\alpha_D$ on the $y$ vs. $m_\varphi$ plane.

 {\it \bf LEP and LHC} 
 Electroweak precision tests at LEP constrain the existence of a new massive photon. In particular, kinetic mixing induces a shift in the mass of the $Z^0$ boson, and the constraint depends on $\epsilon$ and only mildly on $m_{A'}$ . In the limit $m_{A'} \ll m_{Z^0}$, $\epsilon$ is constrained to be $\epsilon < 0.03$ \cite{Hook:2010tw, Curtin:2014cca}.

DM could also give rise to missing energy ($\displaystyle{\not}{E}_T$) signals at the LHC, for example in association with a jet or photon. We recast the results of a CMS DM search \cite{Khachatryan:2014rra}  in the monojet $+\displaystyle{\not}{E}_T$ channel using 20 fb$^{-1}$ at $\sqrt{s}=8\,\TeV$, which bounds $\epsilon \lsim 0.1$.   
The signal yield for this search scales as $\epsilon^2$ and is set by kinematic cuts, whose 
efficiency is roughly flat for the MeV--GeV masses, so the $y$-scaling is similar to that of $B$-factory bounds. 

{\it \bf Beam-Dump Experiments}
Fixed-target beam dump experiments offer another powerful probe of light DM. LSND data sets the strongest constraints for $m_{A'} \lsim m_{\pi^0}$ \cite{Batell:2009di,deNiverville:2011it, deNiverville:2012ij,Dobrescu:2014fca}, via $p N \rightarrow \pi^0+X$, with $\pi^0\rightarrow\gamma A' \to \gamma \bar\chi\chi$, with subsequent DM scattering off an electron, nucleon, or a nucleus mimicking neutrino neutral-current scattering. 
The electron beam-dump experiment E137 is sensitive to $e N \rightarrow e N{\apr},\,{\apr}\rightarrow \bar\chi\chi$ with subsequent DM-electron scattering in a downstream detector \cite{Batell:2014mga}.
In both cases, the production yield scales with couplings as $\epsilon^2$ and the detection probability as $\epsilon^2 \alpha_D$, so that overall yield scales as $\epsilon^4 \alpha_D$.  The benchmark $\alpha_D=1/2$ yields a conservative bound on $y$ that improves as $1 / \sqrt{\alpha_D}$ for smaller $\alpha_D$.  The $m_{A'}$-dependence of these experiments' yields is more complex, but again the choice $m_\varphi / m_{A'}=1 / 3$ is conservative, with stronger constraints for larger ratios.  
%

{\it \bf DM Self-Interactions} 
Constraints from the Bullet Cluster and from cluster lensing \cite{Markevitch:2003at,MiraldaEscude:2000qt} bound the self-interaction cross section to be 
 $\sigma_{\rm self}/m_{\rm DM} \lsim ~\cm^2/\gram$, which implies $\alpha_D \lsim  0.06 (\MeV/m_\chi)^{1/2}  (m_{\apr}/10 \, \MeV)^2  $ and is only relevant at low mass.  Saturating this limit on $\alpha_D$ instead of using our benchmark $\alpha_D=1/2$ leads to a tighter constraint on $y$ from LSND for low ($m_\chi \lsim 10~\MeV$) masses, which is plotted as a green dashed line in Fig.  \ref{fig:money} (again assuming a conservative benchmark $m_\varphi / m_{A'}=1 / 3$). 

{\it \bf Supernovae} 
The production of any free-streaming $\bar\chi\chi$ pairs inside supernovae (SN) is constrained by observations of SN 1987A \cite{Bionta:1987qt,Hirata:1987hu}. 
As discussed in \cite{Izaguirre:2014bca}, the SN core luminosity can be appreciable, but the scattering of $\chi$ off baryons with 
cross section $\sigma =4\pi\alpha \alpha_D\epsilon^2\frac{T_{SN}^2}{m_{A'}^4}$, where $T_{SN}\sim 10$ MeV is the core temperature, 
can be large enough for the $\chi$ diffusion escape time to fall below a few seconds. This occurs for $\alpha_D\epsilon^2 > O({\rm few})\times 10^{-14}$ (or $y\gsim 10^{-16}$ in the above Figures) in the $m_{\apr} \sim m_\chi \sim 10$ MeV range of relevance for these reactions. 
Thus, SN constraints do not probe the thermal dark matter target near $y\gsim 10^{-15}$ at low masses, though a more detailed analysis 
of this physics is called for.

{\bf Visibly Decaying Mediators}
If $m_{\apr}< 2 m_{\varphi(\chi)}$, the $\apr$ signals at colliders are quite rich --- the accelerator-based constraints discussed above have counterparts proceeding through an off-shell $\apr$, and searches for visible decays $\apr \rightarrow \ell^+\ell^-$ become quite powerful.  Numerous experiments \cite{Bjorken:1988as,Riordan:1987aw,Bross:1989mp,Davoudiasl:2012ig,Pospelov:2008zw,Endo:2012hp,Babusci:2012cr,Adlarson:2013eza,Abrahamyan:2011gv,Merkel:2011ze,Bjorken:2009mm,Reece:2009un,Aubert:2009cp,Dent:2012mx,Dreiner:2013mua} constrain this parameter space, and several more are expected to run in the next few years \cite{Essig:2010xa,Battaglieri:2014hga,Freytsis:2009bh,Wojtsekhowski:2009vz,Wojtsekhowski:2012zq,Beranek:2013yqa}.   These results and new experiments are summarized in \cite{Essig:2013lka}.   The interplay of these constraints is illustrated in Figure \ref{fig:visible} for inelastic fermion DM with representative parameter choices $m_{A'} = 1.1 m_\chi$ and $\alpha_D = 0.5$.  The situation is qualitatively similar for the scalar and asymmetric DM scenarios.

%

\section{The Status and Future of Light DM Science}

{\it \bf Present Status}
While scalar mediated light DM scattering is tightly constrained by meson decay, vector mediated scattering is viable over a wide range of mass and couplings. 
Figs.~\ref{fig:money} and \ref{fig:visible} illustrate that current experiments are several orders of magnitude away from decisively testing thermal vector-portal dark matter.  
Electron-scattering limits are 2--3 orders of magnitude short of the thermal target even at the optimal DM masses for these experiments; CMB energy injection bounds exclude the thermal target for symmetric Dirac fermion DM, but are far from constraining pseudo-Dirac, asymmetric, and scalar DM scenarios. 
Low-energy accelerator-based experiments have made significant progress exploring this parameter space -- BaBar data does robustly exclude scalar vector-portal DM with few-GeV mass -- but they too leave 2--4 orders of magnitude of unexplored parameter space below a GeV. 
{\it \bf Future Probes With Existing  Strategies}
Progress searching for light DM based on existing experimental strategies will be driven by three fronts: direct detection, B-factories, and fixed-target experiments.
Direct detection through electron scattering is currently background-limited; future improvements depend on this relatively new field's success in pushing experimental thresholds to lower energies and minimizing backgrounds \cite{Essig:2012yx,Essig:2011nj}. In parallel, nuclear-recoil direct detection experiments are expected to lower their target masses and recoil energy thresholds enough to start exploring the sub-GeV region \cite{Chavarria:2014ika, Cushman:2013zza, Gerbier:2014jwa}. 
Belle II, an upcoming high-luminosity B-factory in Japan, can also significantly improve on current B-factory sensitivity to light DM if it is instrumented with a mono-photon trigger \cite{Essig:2013vha}.  Together, these experiments can explore the viable thermal light DM parameter space above 250--400 MeV.
Fixed-target beam-dump experiments hold significant promise to probe lighter end of this range:  
future neutrino facilities can improve sensitivity for DM below the pion threshold \cite{Kahn:2014sra}, while future electron beam-dumps with relatively small forward detectors extending the sensitivity for $m_\chi\gsim m_{\pi}/2$ \cite{Izaguirre:2013uxa, Izaguirre:2014dua}.  
The fixed-target program to search for visibly decaying mediators, summarized in \cite{Essig:2013lka}, will play an important role in testing the intermediate-mass scenario shown in Figure \ref{fig:visible}.


%

{\it \bf Fixed-Target Missing Momentum Concept}
A particularly powerful probe of vector-portal light DM is to search for missing momentum in electron-nucleus fixed target collisions \cite{Andreas:2013lya,Izaguirre:2014bca}, based on the observation that most of the beam energy in processes $e N \rightarrow eN \apr,\,\apr\rightarrow\bar\chi\chi$ (and similarly for off-shell $A'$) is typically carried by the invisible $\bar\chi\chi$ pair.  
Such experiments' signal rate scales as $\propto \epsilon^2$ (as opposed to $\epsilon^4$ for beam-dump searches). 
The authors have argued in \cite{Izaguirre:2014bca} that such an experiment can suppress backgrounds to $<1$ event in $10^{16}$ electrons on target, provided the detector is capable of measuring the recoiling electron's transverse momentum and vetoing products of rare photo-nuclear reactions.  
Figures~\ref{fig:money} and \ref{fig:visible} illustrate the significant potential of this approach.



{\bf Scenarios Without a Thermal-Relic Milestone}
For DM heavier than the mediator, $t$-channel annihilation into two mediators typically dominates over the $s$-channel processes used to compute the ``relic density'' curve in Figures \ref{fig:money} and \ref{fig:visible}.   For scalar DM and/or vector mediators, this process is never $p$-wave suppressed, and in the inelastic scenario it allows two of the lighter $\chi_-$ particles to annihilate one another --- thus, thermal DM models of these types are inconsistent with the CMB constraint.  However, there are several scenarios compatible with the CMB where annihilation into two mediators dominates: annihilation of fermion DM into scalar mediators ($p$-wave), annihilation into vector mediators \emph{heavier} than the DM (relying on the Boltzmann tail \cite{JoshAndRafaelle}), and asymmetric DM.   The thermal relic abundance cannot be used to predict a lower-bound on $y$ in these cases.  A more appropriate figure of merit than $y$ for comparing direct detection and accelerator-based experiments for $m_{\apr} < m_{\varphi(\chi)}$ is $\alpha \alpha_D \epsilon^2$ (or similarly for scalar mediators), where the direct detection limits are evaluated conservatively for $m_{\apr} \approx m_{\varphi(\chi)}$.  

{\bf Summary}
This article shows that many of the simplest models for sub-GeV thermal DM are consistent with all current data.  A generic and simple possibility is that DM couples to the SM through a kinetically mixed dark photon. Although symmetric Dirac fermion DM annihilating through the vector portal is excluded by CMB limits, scalar, pseudo-Dirac, and asymmetric DM scenarios are all largely untested.  These scenarios define sharp milestones for future experiments to reach.  
Together, planned $B$-factories (Belle-II, if equipped with a mono-photon trigger), direct detection experiments (SuperCDMS), and possible electron fixed-target experiments based on missing momentum should be capable of reaching this target over almost all of the MeV-to-GeV mass range.

\begin{acknowledgments}
{\it Acknowledgments}  We thank Wolfgang Altmanshoffer, Mariana Gonzalez, Valentin Hirshi, Yoni Kahn, David Morrissey, Maxim Pospelov, Joshua Ruderman, Tracy Slatyer, and Kathryn Zurek for helpful conversations.  This research was supported in part by Perimeter Institute for Theoretical Physics. Research at Perimeter Institute is supported by the Government of Canada through Industry Canada and by the Province of Ontario through the Ministry of Research and Innovation.
\end{acknowledgments} \medskip


\bigskip

\bibliography{mmprl}

\begin{thebibliography}{73}%
\makeatletter
\providecommand \@ifxundefined [1]{%
 \@ifx{#1\undefined}
}%
\providecommand \@ifnum [1]{%
 \ifnum #1\expandafter \@firstoftwo
 \else \expandafter \@secondoftwo
 \fi
}%
\providecommand \@ifx [1]{%
 \ifx #1\expandafter \@firstoftwo
 \else \expandafter \@secondoftwo
 \fi
}%
\providecommand \natexlab [1]{#1}%
\providecommand \enquote  [1]{``#1''}%
\providecommand \bibnamefont  [1]{#1}%
\providecommand \bibfnamefont [1]{#1}%
\providecommand \citenamefont [1]{#1}%
\providecommand \href@noop [0]{\@secondoftwo}%
\providecommand \href [0]{\begingroup \@sanitize@url \@href}%
\providecommand \@href[1]{\@@startlink{#1}\@@href}%
\providecommand \@@href[1]{\endgroup#1\@@endlink}%
\providecommand \@sanitize@url [0]{\catcode `\\12\catcode `\$12\catcode
  `\&12\catcode `\#12\catcode `\^12\catcode `\_12\catcode `\%12\relax}%
\providecommand \@@startlink[1]{}%
\providecommand \@@endlink[0]{}%
\providecommand \url  [0]{\begingroup\@sanitize@url \@url }%
\providecommand \@url [1]{\endgroup\@href {#1}{\urlprefix }}%
\providecommand \urlprefix  [0]{URL }%
\providecommand \Eprint [0]{\href }%
\providecommand \doibase [0]{http://dx.doi.org/}%
\providecommand \selectlanguage [0]{\@gobble}%
\providecommand \bibinfo  [0]{\@secondoftwo}%
\providecommand \bibfield  [0]{\@secondoftwo}%
\providecommand \translation [1]{[#1]}%
\providecommand \BibitemOpen [0]{}%
\providecommand \bibitemStop [0]{}%
\providecommand \bibitemNoStop [0]{.\EOS\space}%
\providecommand \EOS [0]{\spacefactor3000\relax}%
\providecommand \BibitemShut  [1]{\csname bibitem#1\endcsname}%
\let\auto@bib@innerbib\@empty
\bibitem [{\citenamefont {Lee}\ and\ \citenamefont
  {Weinberg}(1977)}]{Lee:1977ua}%
  \BibitemOpen
  \bibfield  {author} {\bibinfo {author} {\bibfnamefont {B.~W.}\ \bibnamefont
  {Lee}}\ and\ \bibinfo {author} {\bibfnamefont {S.}~\bibnamefont {Weinberg}},\
  }\href {\doibase 10.1103/PhysRevLett.39.165} {\bibfield  {journal} {\bibinfo
  {journal} {Phys.Rev.Lett.}\ }\textbf {\bibinfo {volume} {39}},\ \bibinfo
  {pages} {165} (\bibinfo {year} {1977})}\BibitemShut {NoStop}%
\bibitem [{\citenamefont {Boehm}\ \emph {et~al.}(2004)\citenamefont {Boehm},
  \citenamefont {Ensslin},\ and\ \citenamefont {Silk}}]{Boehm:2002yz}%
  \BibitemOpen
  \bibfield  {author} {\bibinfo {author} {\bibfnamefont {C.}~\bibnamefont
  {Boehm}}, \bibinfo {author} {\bibfnamefont {T.}~\bibnamefont {Ensslin}}, \
  and\ \bibinfo {author} {\bibfnamefont {J.}~\bibnamefont {Silk}},\ }\href
  {\doibase 10.1088/0954-3899/30/3/004} {\bibfield  {journal} {\bibinfo
  {journal} {J.Phys.}\ }\textbf {\bibinfo {volume} {G30}},\ \bibinfo {pages}
  {279} (\bibinfo {year} {2004})},\ \Eprint
  {http://arxiv.org/abs/astro-ph/0208458} {arXiv:astro-ph/0208458 [astro-ph]}
  \BibitemShut {NoStop}%
\bibitem [{\citenamefont {Boehm}\ and\ \citenamefont
  {Fayet}(2004)}]{Boehm:2003hm}%
  \BibitemOpen
  \bibfield  {author} {\bibinfo {author} {\bibfnamefont {C.}~\bibnamefont
  {Boehm}}\ and\ \bibinfo {author} {\bibfnamefont {P.}~\bibnamefont {Fayet}},\
  }\href {\doibase 10.1016/j.nuclphysb.2004.01.015} {\bibfield  {journal}
  {\bibinfo  {journal} {Nucl.Phys.}\ }\textbf {\bibinfo {volume} {B683}},\
  \bibinfo {pages} {219} (\bibinfo {year} {2004})},\ \Eprint
  {http://arxiv.org/abs/hep-ph/0305261} {arXiv:hep-ph/0305261 [hep-ph]}
  \BibitemShut {NoStop}%
\bibitem [{\citenamefont {Pospelov}\ \emph {et~al.}(2008)\citenamefont
  {Pospelov}, \citenamefont {Ritz},\ and\ \citenamefont
  {Voloshin}}]{Pospelov:2007mp}%
  \BibitemOpen
  \bibfield  {author} {\bibinfo {author} {\bibfnamefont {M.}~\bibnamefont
  {Pospelov}}, \bibinfo {author} {\bibfnamefont {A.}~\bibnamefont {Ritz}}, \
  and\ \bibinfo {author} {\bibfnamefont {M.~B.}\ \bibnamefont {Voloshin}},\
  }\href {\doibase 10.1016/j.physletb.2008.02.052} {\bibfield  {journal}
  {\bibinfo  {journal} {Phys.Lett.}\ }\textbf {\bibinfo {volume} {B662}},\
  \bibinfo {pages} {53} (\bibinfo {year} {2008})},\ \Eprint
  {http://arxiv.org/abs/0711.4866} {arXiv:0711.4866 [hep-ph]} \BibitemShut
  {NoStop}%
\bibitem [{\citenamefont {Pospelov}(2009)}]{Pospelov:2008zw}%
  \BibitemOpen
  \bibfield  {author} {\bibinfo {author} {\bibfnamefont {M.}~\bibnamefont
  {Pospelov}},\ }\href {\doibase 10.1103/PhysRevD.80.095002} {\bibfield
  {journal} {\bibinfo  {journal} {Phys.Rev.}\ }\textbf {\bibinfo {volume}
  {D80}},\ \bibinfo {pages} {095002} (\bibinfo {year} {2009})},\ \Eprint
  {http://arxiv.org/abs/0811.1030} {arXiv:0811.1030 [hep-ph]} \BibitemShut
  {NoStop}%
\bibitem [{\citenamefont {Arkani-Hamed}\ \emph {et~al.}(2009)\citenamefont
  {Arkani-Hamed}, \citenamefont {Finkbeiner}, \citenamefont {Slatyer},\ and\
  \citenamefont {Weiner}}]{ArkaniHamed:2008qn}%
  \BibitemOpen
  \bibfield  {author} {\bibinfo {author} {\bibfnamefont {N.}~\bibnamefont
  {Arkani-Hamed}}, \bibinfo {author} {\bibfnamefont {D.~P.}\ \bibnamefont
  {Finkbeiner}}, \bibinfo {author} {\bibfnamefont {T.~R.}\ \bibnamefont
  {Slatyer}}, \ and\ \bibinfo {author} {\bibfnamefont {N.}~\bibnamefont
  {Weiner}},\ }\href {\doibase 10.1103/PhysRevD.79.015014} {\bibfield
  {journal} {\bibinfo  {journal} {Phys.Rev.}\ }\textbf {\bibinfo {volume}
  {D79}},\ \bibinfo {pages} {015014} (\bibinfo {year} {2009})},\ \Eprint
  {http://arxiv.org/abs/0810.0713} {arXiv:0810.0713 [hep-ph]} \BibitemShut
  {NoStop}%
\bibitem [{\citenamefont {Pospelov}\ and\ \citenamefont
  {Ritz}(2009)}]{secluded-wimps}%
  \BibitemOpen
  \bibfield  {author} {\bibinfo {author} {\bibfnamefont {M.}~\bibnamefont
  {Pospelov}}\ and\ \bibinfo {author} {\bibfnamefont {A.}~\bibnamefont
  {Ritz}},\ }\href {\doibase 10.1016/j.physletb.2008.12.012} {\bibfield
  {journal} {\bibinfo  {journal} {Phys.Lett.}\ }\textbf {\bibinfo {volume}
  {B671}},\ \bibinfo {pages} {391} (\bibinfo {year} {2009})},\ \Eprint
  {http://arxiv.org/abs/0810.1502} {arXiv:0810.1502 [hep-ph]} \BibitemShut
  {NoStop}%
\bibitem [{\citenamefont {Okun}(1982)}]{Okun:1982xi}%
  \BibitemOpen
  \bibfield  {author} {\bibinfo {author} {\bibfnamefont {L.}~\bibnamefont
  {Okun}},\ }\href@noop {} {\bibfield  {journal} {\bibinfo  {journal}
  {Sov.Phys.JETP}\ }\textbf {\bibinfo {volume} {56}},\ \bibinfo {pages} {502}
  (\bibinfo {year} {1982})}\BibitemShut {NoStop}%
\bibitem [{\citenamefont {Holdom}(1986)}]{Holdom:1985ag}%
  \BibitemOpen
  \bibfield  {author} {\bibinfo {author} {\bibfnamefont {B.}~\bibnamefont
  {Holdom}},\ }\href {\doibase 10.1016/0370-2693(86)91377-8} {\bibfield
  {journal} {\bibinfo  {journal} {Phys.Lett.}\ }\textbf {\bibinfo {volume}
  {B166}},\ \bibinfo {pages} {196} (\bibinfo {year} {1986})}\BibitemShut
  {NoStop}%
\bibitem [{\citenamefont {Izaguirre}\ \emph {et~al.}()\citenamefont
  {Izaguirre}, \citenamefont {Krnjaic}, \citenamefont {Schuster},\ and\
  \citenamefont {Toro}}]{OurFuturePaper}%
  \BibitemOpen
  \bibfield  {author} {\bibinfo {author} {\bibfnamefont {E.}~\bibnamefont
  {Izaguirre}}, \bibinfo {author} {\bibfnamefont {G.}~\bibnamefont {Krnjaic}},
  \bibinfo {author} {\bibfnamefont {P.}~\bibnamefont {Schuster}}, \ and\
  \bibinfo {author} {\bibfnamefont {N.}~\bibnamefont {Toro}},\ }\href@noop {}
  {\ }\Eprint {http://arxiv.org/abs/{\it in preparation}} {{\it in
  preparation}} \BibitemShut {NoStop}%
\bibitem [{\citenamefont {Tucker-Smith}\ and\ \citenamefont
  {Weiner}(2001)}]{TuckerSmith:2001hy}%
  \BibitemOpen
  \bibfield  {author} {\bibinfo {author} {\bibfnamefont {D.}~\bibnamefont
  {Tucker-Smith}}\ and\ \bibinfo {author} {\bibfnamefont {N.}~\bibnamefont
  {Weiner}},\ }\href {\doibase 10.1103/PhysRevD.64.043502} {\bibfield
  {journal} {\bibinfo  {journal} {Phys.Rev.}\ }\textbf {\bibinfo {volume}
  {D64}},\ \bibinfo {pages} {043502} (\bibinfo {year} {2001})},\ \Eprint
  {http://arxiv.org/abs/hep-ph/0101138} {arXiv:hep-ph/0101138 [hep-ph]}
  \BibitemShut {NoStop}%
\bibitem [{\citenamefont {Altmannshofer}\ \emph
  {et~al.}(2014{\natexlab{a}})\citenamefont {Altmannshofer}, \citenamefont
  {Gori}, \citenamefont {Pospelov},\ and\ \citenamefont
  {Yavin}}]{Altmannshofer:2014cfa}%
  \BibitemOpen
  \bibfield  {author} {\bibinfo {author} {\bibfnamefont {W.}~\bibnamefont
  {Altmannshofer}}, \bibinfo {author} {\bibfnamefont {S.}~\bibnamefont {Gori}},
  \bibinfo {author} {\bibfnamefont {M.}~\bibnamefont {Pospelov}}, \ and\
  \bibinfo {author} {\bibfnamefont {I.}~\bibnamefont {Yavin}},\ }\href
  {\doibase 10.1103/PhysRevD.89.095033} {\bibfield  {journal} {\bibinfo
  {journal} {Phys.Rev.}\ }\textbf {\bibinfo {volume} {D89}},\ \bibinfo {pages}
  {095033} (\bibinfo {year} {2014}{\natexlab{a}})},\ \Eprint
  {http://arxiv.org/abs/1403.1269} {arXiv:1403.1269 [hep-ph]} \BibitemShut
  {NoStop}%
\bibitem [{\citenamefont {Altmannshofer}\ \emph
  {et~al.}(2014{\natexlab{b}})\citenamefont {Altmannshofer}, \citenamefont
  {Gori}, \citenamefont {Pospelov},\ and\ \citenamefont
  {Yavin}}]{Altmannshofer:2014pba}%
  \BibitemOpen
  \bibfield  {author} {\bibinfo {author} {\bibfnamefont {W.}~\bibnamefont
  {Altmannshofer}}, \bibinfo {author} {\bibfnamefont {S.}~\bibnamefont {Gori}},
  \bibinfo {author} {\bibfnamefont {M.}~\bibnamefont {Pospelov}}, \ and\
  \bibinfo {author} {\bibfnamefont {I.}~\bibnamefont {Yavin}},\ }\href
  {\doibase 10.1103/PhysRevLett.113.091801} {\bibfield  {journal} {\bibinfo
  {journal} {Phys.Rev.Lett.}\ }\textbf {\bibinfo {volume} {113}},\ \bibinfo
  {pages} {091801} (\bibinfo {year} {2014}{\natexlab{b}})},\ \Eprint
  {http://arxiv.org/abs/1406.2332} {arXiv:1406.2332 [hep-ph]} \BibitemShut
  {NoStop}%
\bibitem [{\citenamefont {Dolan}\ \emph {et~al.}(2014)\citenamefont {Dolan},
  \citenamefont {McCabe}, \citenamefont {Kahlhoefer},\ and\ \citenamefont
  {Schmidt-Hoberg}}]{Dolan:2014ska}%
  \BibitemOpen
  \bibfield  {author} {\bibinfo {author} {\bibfnamefont {M.~J.}\ \bibnamefont
  {Dolan}}, \bibinfo {author} {\bibfnamefont {C.}~\bibnamefont {McCabe}},
  \bibinfo {author} {\bibfnamefont {F.}~\bibnamefont {Kahlhoefer}}, \ and\
  \bibinfo {author} {\bibfnamefont {K.}~\bibnamefont {Schmidt-Hoberg}},\
  }\href@noop {} {\  (\bibinfo {year} {2014})},\ \Eprint
  {http://arxiv.org/abs/1412.5174} {arXiv:1412.5174 [hep-ph]} \BibitemShut
  {NoStop}%
\bibitem [{\citenamefont {Bird}\ \emph {et~al.}(2004)\citenamefont {Bird},
  \citenamefont {Jackson}, \citenamefont {Kowalewski},\ and\ \citenamefont
  {Pospelov}}]{Bird:2004ts}%
  \BibitemOpen
  \bibfield  {author} {\bibinfo {author} {\bibfnamefont {C.}~\bibnamefont
  {Bird}}, \bibinfo {author} {\bibfnamefont {P.}~\bibnamefont {Jackson}},
  \bibinfo {author} {\bibfnamefont {R.~V.}\ \bibnamefont {Kowalewski}}, \ and\
  \bibinfo {author} {\bibfnamefont {M.}~\bibnamefont {Pospelov}},\ }\href
  {\doibase 10.1103/PhysRevLett.93.201803} {\bibfield  {journal} {\bibinfo
  {journal} {Phys.Rev.Lett.}\ }\textbf {\bibinfo {volume} {93}},\ \bibinfo
  {pages} {201803} (\bibinfo {year} {2004})},\ \Eprint
  {http://arxiv.org/abs/hep-ph/0401195} {arXiv:hep-ph/0401195 [hep-ph]}
  \BibitemShut {NoStop}%
\bibitem [{\citenamefont {Lees}\ \emph {et~al.}(2013)\citenamefont {Lees} \emph
  {et~al.}}]{Lees:2013kla}%
  \BibitemOpen
  \bibfield  {author} {\bibinfo {author} {\bibfnamefont {J.}~\bibnamefont
  {Lees}} \emph {et~al.} (\bibinfo {collaboration} {BaBar}),\ }\href {\doibase
  10.1103/PhysRevD.87.112005} {\bibfield  {journal} {\bibinfo  {journal}
  {Phys.Rev.}\ }\textbf {\bibinfo {volume} {D87}},\ \bibinfo {pages} {112005}
  (\bibinfo {year} {2013})},\ \Eprint {http://arxiv.org/abs/1303.7465}
  {arXiv:1303.7465 [hep-ex]} \BibitemShut {NoStop}%
\bibitem [{\citenamefont {Finkbeiner}\ \emph {et~al.}(2012)\citenamefont
  {Finkbeiner}, \citenamefont {Galli}, \citenamefont {Lin},\ and\ \citenamefont
  {Slatyer}}]{Finkbeiner:2011dx}%
  \BibitemOpen
  \bibfield  {author} {\bibinfo {author} {\bibfnamefont {D.~P.}\ \bibnamefont
  {Finkbeiner}}, \bibinfo {author} {\bibfnamefont {S.}~\bibnamefont {Galli}},
  \bibinfo {author} {\bibfnamefont {T.}~\bibnamefont {Lin}}, \ and\ \bibinfo
  {author} {\bibfnamefont {T.~R.}\ \bibnamefont {Slatyer}},\ }\href {\doibase
  10.1103/PhysRevD.85.043522} {\bibfield  {journal} {\bibinfo  {journal}
  {Phys.Rev.}\ }\textbf {\bibinfo {volume} {D85}},\ \bibinfo {pages} {043522}
  (\bibinfo {year} {2012})},\ \Eprint {http://arxiv.org/abs/1109.6322}
  {arXiv:1109.6322 [astro-ph.CO]} \BibitemShut {NoStop}%
\bibitem [{\citenamefont {Lin}\ \emph {et~al.}(2012)\citenamefont {Lin},
  \citenamefont {Yu},\ and\ \citenamefont {Zurek}}]{Lin:2011gj}%
  \BibitemOpen
  \bibfield  {author} {\bibinfo {author} {\bibfnamefont {T.}~\bibnamefont
  {Lin}}, \bibinfo {author} {\bibfnamefont {H.-B.}\ \bibnamefont {Yu}}, \ and\
  \bibinfo {author} {\bibfnamefont {K.~M.}\ \bibnamefont {Zurek}},\ }\href
  {\doibase 10.1103/PhysRevD.85.063503} {\bibfield  {journal} {\bibinfo
  {journal} {Phys.Rev.}\ }\textbf {\bibinfo {volume} {D85}},\ \bibinfo {pages}
  {063503} (\bibinfo {year} {2012})},\ \Eprint {http://arxiv.org/abs/1111.0293}
  {arXiv:1111.0293 [hep-ph]} \BibitemShut {NoStop}%
\bibitem [{\citenamefont {Galli}\ \emph {et~al.}(2011)\citenamefont {Galli},
  \citenamefont {Iocco}, \citenamefont {Bertone},\ and\ \citenamefont
  {Melchiorri}}]{Galli:2011rz}%
  \BibitemOpen
  \bibfield  {author} {\bibinfo {author} {\bibfnamefont {S.}~\bibnamefont
  {Galli}}, \bibinfo {author} {\bibfnamefont {F.}~\bibnamefont {Iocco}},
  \bibinfo {author} {\bibfnamefont {G.}~\bibnamefont {Bertone}}, \ and\
  \bibinfo {author} {\bibfnamefont {A.}~\bibnamefont {Melchiorri}},\ }\href
  {\doibase 10.1103/PhysRevD.84.027302} {\bibfield  {journal} {\bibinfo
  {journal} {Phys.Rev.}\ }\textbf {\bibinfo {volume} {D84}},\ \bibinfo {pages}
  {027302} (\bibinfo {year} {2011})},\ \Eprint {http://arxiv.org/abs/1106.1528}
  {arXiv:1106.1528 [astro-ph.CO]} \BibitemShut {NoStop}%
\bibitem [{\citenamefont {Madhavacheril}\ \emph {et~al.}(2014)\citenamefont
  {Madhavacheril}, \citenamefont {Sehgal},\ and\ \citenamefont
  {Slatyer}}]{Madhavacheril:2013cna}%
  \BibitemOpen
  \bibfield  {author} {\bibinfo {author} {\bibfnamefont {M.~S.}\ \bibnamefont
  {Madhavacheril}}, \bibinfo {author} {\bibfnamefont {N.}~\bibnamefont
  {Sehgal}}, \ and\ \bibinfo {author} {\bibfnamefont {T.~R.}\ \bibnamefont
  {Slatyer}},\ }\href {\doibase 10.1103/PhysRevD.89.103508} {\bibfield
  {journal} {\bibinfo  {journal} {Phys.Rev.}\ }\textbf {\bibinfo {volume}
  {D89}},\ \bibinfo {pages} {103508} (\bibinfo {year} {2014})},\ \Eprint
  {http://arxiv.org/abs/1310.3815} {arXiv:1310.3815 [astro-ph.CO]} \BibitemShut
  {NoStop}%
\bibitem [{\citenamefont {Ade}\ \emph {et~al.}(2014)\citenamefont {Ade} \emph
  {et~al.}}]{Ade:2013zuv}%
  \BibitemOpen
  \bibfield  {author} {\bibinfo {author} {\bibfnamefont {P.}~\bibnamefont
  {Ade}} \emph {et~al.} (\bibinfo {collaboration} {Planck}),\ }\href {\doibase
  10.1051/0004-6361/201321591} {\bibfield  {journal} {\bibinfo  {journal}
  {Astron.Astrophys.}\ }\textbf {\bibinfo {volume} {571}},\ \bibinfo {pages}
  {A16} (\bibinfo {year} {2014})},\ \Eprint {http://arxiv.org/abs/1303.5076}
  {arXiv:1303.5076 [astro-ph.CO]} \BibitemShut {NoStop}%
\bibitem [{\citenamefont {Gonzalez}\ and\ \citenamefont
  {Toro}()}]{IDMcosmology}%
  \BibitemOpen
  \bibfield  {author} {\bibinfo {author} {\bibfnamefont {M.}~\bibnamefont
  {Gonzalez}}\ and\ \bibinfo {author} {\bibfnamefont {N.}~\bibnamefont
  {Toro}},\ }\href@noop {} {\ }\Eprint {http://arxiv.org/abs/{\it in
  preparation}} {{\it in preparation}} \BibitemShut {NoStop}%
\bibitem [{\citenamefont {Hisano}\ \emph
  {et~al.}(2009{\natexlab{a}})\citenamefont {Hisano}, \citenamefont {Kawasaki},
  \citenamefont {Kohri},\ and\ \citenamefont {Nakayama}}]{Hisano:2008ti}%
  \BibitemOpen
  \bibfield  {author} {\bibinfo {author} {\bibfnamefont {J.}~\bibnamefont
  {Hisano}}, \bibinfo {author} {\bibfnamefont {M.}~\bibnamefont {Kawasaki}},
  \bibinfo {author} {\bibfnamefont {K.}~\bibnamefont {Kohri}}, \ and\ \bibinfo
  {author} {\bibfnamefont {K.}~\bibnamefont {Nakayama}},\ }\href {\doibase
  10.1103/PhysRevD.79.063514, 10.1103/PhysRevD.80.029907,
  10.1103/PhysRevD.79.063514 10.1103/PhysRevD.80.029907} {\bibfield  {journal}
  {\bibinfo  {journal} {Phys.Rev.}\ }\textbf {\bibinfo {volume} {D79}},\
  \bibinfo {pages} {063514} (\bibinfo {year} {2009}{\natexlab{a}})},\ \Eprint
  {http://arxiv.org/abs/0810.1892} {arXiv:0810.1892 [hep-ph]} \BibitemShut
  {NoStop}%
\bibitem [{\citenamefont {Hisano}\ \emph
  {et~al.}(2009{\natexlab{b}})\citenamefont {Hisano}, \citenamefont {Kawasaki},
  \citenamefont {Kohri}, \citenamefont {Moroi},\ and\ \citenamefont
  {Nakayama}}]{Hisano:2009rc}%
  \BibitemOpen
  \bibfield  {author} {\bibinfo {author} {\bibfnamefont {J.}~\bibnamefont
  {Hisano}}, \bibinfo {author} {\bibfnamefont {M.}~\bibnamefont {Kawasaki}},
  \bibinfo {author} {\bibfnamefont {K.}~\bibnamefont {Kohri}}, \bibinfo
  {author} {\bibfnamefont {T.}~\bibnamefont {Moroi}}, \ and\ \bibinfo {author}
  {\bibfnamefont {K.}~\bibnamefont {Nakayama}},\ }\href {\doibase
  10.1103/PhysRevD.79.083522} {\bibfield  {journal} {\bibinfo  {journal}
  {Phys.Rev.}\ }\textbf {\bibinfo {volume} {D79}},\ \bibinfo {pages} {083522}
  (\bibinfo {year} {2009}{\natexlab{b}})},\ \Eprint
  {http://arxiv.org/abs/0901.3582} {arXiv:0901.3582 [hep-ph]} \BibitemShut
  {NoStop}%
\bibitem [{\citenamefont {Jedamzik}\ and\ \citenamefont
  {Pospelov}(2009)}]{Jedamzik:2009uy}%
  \BibitemOpen
  \bibfield  {author} {\bibinfo {author} {\bibfnamefont {K.}~\bibnamefont
  {Jedamzik}}\ and\ \bibinfo {author} {\bibfnamefont {M.}~\bibnamefont
  {Pospelov}},\ }\href {\doibase 10.1088/1367-2630/11/10/105028} {\bibfield
  {journal} {\bibinfo  {journal} {New J.Phys.}\ }\textbf {\bibinfo {volume}
  {11}},\ \bibinfo {pages} {105028} (\bibinfo {year} {2009})},\ \Eprint
  {http://arxiv.org/abs/0906.2087} {arXiv:0906.2087 [hep-ph]} \BibitemShut
  {NoStop}%
\bibitem [{\citenamefont {Henning}\ and\ \citenamefont
  {Murayama}(2012)}]{Henning:2012rm}%
  \BibitemOpen
  \bibfield  {author} {\bibinfo {author} {\bibfnamefont {B.}~\bibnamefont
  {Henning}}\ and\ \bibinfo {author} {\bibfnamefont {H.}~\bibnamefont
  {Murayama}},\ }\href@noop {} {\  (\bibinfo {year} {2012})},\ \Eprint
  {http://arxiv.org/abs/1205.6479} {arXiv:1205.6479 [hep-ph]} \BibitemShut
  {NoStop}%
\bibitem [{\citenamefont {Essig}\ \emph
  {et~al.}(2012{\natexlab{a}})\citenamefont {Essig}, \citenamefont
  {Manalaysay}, \citenamefont {Mardon}, \citenamefont {Sorensen},\ and\
  \citenamefont {Volansky}}]{Essig:2012yx}%
  \BibitemOpen
  \bibfield  {author} {\bibinfo {author} {\bibfnamefont {R.}~\bibnamefont
  {Essig}}, \bibinfo {author} {\bibfnamefont {A.}~\bibnamefont {Manalaysay}},
  \bibinfo {author} {\bibfnamefont {J.}~\bibnamefont {Mardon}}, \bibinfo
  {author} {\bibfnamefont {P.}~\bibnamefont {Sorensen}}, \ and\ \bibinfo
  {author} {\bibfnamefont {T.}~\bibnamefont {Volansky}},\ }\href {\doibase
  10.1103/PhysRevLett.109.021301} {\bibfield  {journal} {\bibinfo  {journal}
  {Phys.Rev.Lett.}\ }\textbf {\bibinfo {volume} {109}},\ \bibinfo {pages}
  {021301} (\bibinfo {year} {2012}{\natexlab{a}})},\ \Eprint
  {http://arxiv.org/abs/1206.2644} {arXiv:1206.2644 [astro-ph.CO]} \BibitemShut
  {NoStop}%
\bibitem [{\citenamefont {Essig}\ \emph
  {et~al.}(2012{\natexlab{b}})\citenamefont {Essig}, \citenamefont {Mardon},\
  and\ \citenamefont {Volansky}}]{Essig:2011nj}%
  \BibitemOpen
  \bibfield  {author} {\bibinfo {author} {\bibfnamefont {R.}~\bibnamefont
  {Essig}}, \bibinfo {author} {\bibfnamefont {J.}~\bibnamefont {Mardon}}, \
  and\ \bibinfo {author} {\bibfnamefont {T.}~\bibnamefont {Volansky}},\ }\href
  {\doibase 10.1103/PhysRevD.85.076007} {\bibfield  {journal} {\bibinfo
  {journal} {Phys.Rev.}\ }\textbf {\bibinfo {volume} {D85}},\ \bibinfo {pages}
  {076007} (\bibinfo {year} {2012}{\natexlab{b}})},\ \Eprint
  {http://arxiv.org/abs/1108.5383} {arXiv:1108.5383 [hep-ph]} \BibitemShut
  {NoStop}%
\bibitem [{\citenamefont {Essig}\ \emph
  {et~al.}(2013{\natexlab{a}})\citenamefont {Essig}, \citenamefont {Jaros},
  \citenamefont {Wester}, \citenamefont {Adrian}, \citenamefont {Andreas} \emph
  {et~al.}}]{Essig:2013lka}%
  \BibitemOpen
  \bibfield  {author} {\bibinfo {author} {\bibfnamefont {R.}~\bibnamefont
  {Essig}}, \bibinfo {author} {\bibfnamefont {J.~A.}\ \bibnamefont {Jaros}},
  \bibinfo {author} {\bibfnamefont {W.}~\bibnamefont {Wester}}, \bibinfo
  {author} {\bibfnamefont {P.~H.}\ \bibnamefont {Adrian}}, \bibinfo {author}
  {\bibfnamefont {S.}~\bibnamefont {Andreas}},  \emph {et~al.},\ }\href@noop {}
  {\  (\bibinfo {year} {2013}{\natexlab{a}})},\ \Eprint
  {http://arxiv.org/abs/1311.0029} {arXiv:1311.0029 [hep-ph]} \BibitemShut
  {NoStop}%
\bibitem [{\citenamefont {D'Agnolo}\ and\ \citenamefont
  {Ruderman}()}]{JoshAndRafaelle}%
  \BibitemOpen
  \bibfield  {author} {\bibinfo {author} {\bibfnamefont {R.~T.}\ \bibnamefont
  {D'Agnolo}}\ and\ \bibinfo {author} {\bibfnamefont {J.}~\bibnamefont
  {Ruderman}},\ }\href@noop {} {\ }\Eprint {http://arxiv.org/abs/{\it in
  preparation}} {{\it in preparation}} \BibitemShut {NoStop}%
\bibitem [{\citenamefont {Izaguirre}\ \emph {et~al.}(2013)\citenamefont
  {Izaguirre}, \citenamefont {Krnjaic}, \citenamefont {Schuster},\ and\
  \citenamefont {Toro}}]{Izaguirre:2013uxa}%
  \BibitemOpen
  \bibfield  {author} {\bibinfo {author} {\bibfnamefont {E.}~\bibnamefont
  {Izaguirre}}, \bibinfo {author} {\bibfnamefont {G.}~\bibnamefont {Krnjaic}},
  \bibinfo {author} {\bibfnamefont {P.}~\bibnamefont {Schuster}}, \ and\
  \bibinfo {author} {\bibfnamefont {N.}~\bibnamefont {Toro}},\ }\href {\doibase
  10.1103/PhysRevD.88.114015} {\bibfield  {journal} {\bibinfo  {journal}
  {Phys.Rev.}\ }\textbf {\bibinfo {volume} {D88}},\ \bibinfo {pages} {114015}
  (\bibinfo {year} {2013})},\ \Eprint {http://arxiv.org/abs/1307.6554}
  {arXiv:1307.6554 [hep-ph]} \BibitemShut {NoStop}%
\bibitem [{\citenamefont {Essig}\ \emph
  {et~al.}(2013{\natexlab{b}})\citenamefont {Essig}, \citenamefont {Mardon},
  \citenamefont {Papucci}, \citenamefont {Volansky},\ and\ \citenamefont
  {Zhong}}]{Essig:2013vha}%
  \BibitemOpen
  \bibfield  {author} {\bibinfo {author} {\bibfnamefont {R.}~\bibnamefont
  {Essig}}, \bibinfo {author} {\bibfnamefont {J.}~\bibnamefont {Mardon}},
  \bibinfo {author} {\bibfnamefont {M.}~\bibnamefont {Papucci}}, \bibinfo
  {author} {\bibfnamefont {T.}~\bibnamefont {Volansky}}, \ and\ \bibinfo
  {author} {\bibfnamefont {Y.-M.}\ \bibnamefont {Zhong}},\ }\href {\doibase
  10.1007/JHEP11(2013)167} {\bibfield  {journal} {\bibinfo  {journal} {JHEP}\
  }\textbf {\bibinfo {volume} {1311}},\ \bibinfo {pages} {167} (\bibinfo {year}
  {2013}{\natexlab{b}})},\ \Eprint {http://arxiv.org/abs/1309.5084}
  {arXiv:1309.5084 [hep-ph]} \BibitemShut {NoStop}%
\bibitem [{\citenamefont {Aubert}\ \emph {et~al.}(2008)\citenamefont {Aubert}
  \emph {et~al.}}]{Aubert:2008as}%
  \BibitemOpen
  \bibfield  {author} {\bibinfo {author} {\bibfnamefont {B.}~\bibnamefont
  {Aubert}} \emph {et~al.} (\bibinfo {collaboration} {BaBar Collaboration}),\
  }\href@noop {} {\  (\bibinfo {year} {2008})},\ \Eprint
  {http://arxiv.org/abs/0808.0017} {arXiv:0808.0017 [hep-ex]} \BibitemShut
  {NoStop}%
\bibitem [{\citenamefont {Davoudiasl}\ and\ \citenamefont
  {Marciano}(2015)}]{Davoudiasl:2015hxa}%
  \BibitemOpen
  \bibfield  {author} {\bibinfo {author} {\bibfnamefont {H.}~\bibnamefont
  {Davoudiasl}}\ and\ \bibinfo {author} {\bibfnamefont {W.~J.}\ \bibnamefont
  {Marciano}},\ }\href@noop {} {\  (\bibinfo {year} {2015})},\ \Eprint
  {http://arxiv.org/abs/1502.07383} {arXiv:1502.07383 [hep-ph]} \BibitemShut
  {NoStop}%
\bibitem [{\citenamefont {Hook}\ \emph {et~al.}(2011)\citenamefont {Hook},
  \citenamefont {Izaguirre},\ and\ \citenamefont {Wacker}}]{Hook:2010tw}%
  \BibitemOpen
  \bibfield  {author} {\bibinfo {author} {\bibfnamefont {A.}~\bibnamefont
  {Hook}}, \bibinfo {author} {\bibfnamefont {E.}~\bibnamefont {Izaguirre}}, \
  and\ \bibinfo {author} {\bibfnamefont {J.~G.}\ \bibnamefont {Wacker}},\
  }\href {\doibase 10.1155/2011/859762} {\bibfield  {journal} {\bibinfo
  {journal} {Adv.High Energy Phys.}\ }\textbf {\bibinfo {volume} {2011}},\
  \bibinfo {pages} {859762} (\bibinfo {year} {2011})},\ \Eprint
  {http://arxiv.org/abs/1006.0973} {arXiv:1006.0973 [hep-ph]} \BibitemShut
  {NoStop}%
\bibitem [{\citenamefont {Curtin}\ \emph {et~al.}(2015)\citenamefont {Curtin},
  \citenamefont {Essig}, \citenamefont {Gori},\ and\ \citenamefont
  {Shelton}}]{Curtin:2014cca}%
  \BibitemOpen
  \bibfield  {author} {\bibinfo {author} {\bibfnamefont {D.}~\bibnamefont
  {Curtin}}, \bibinfo {author} {\bibfnamefont {R.}~\bibnamefont {Essig}},
  \bibinfo {author} {\bibfnamefont {S.}~\bibnamefont {Gori}}, \ and\ \bibinfo
  {author} {\bibfnamefont {J.}~\bibnamefont {Shelton}},\ }\href {\doibase
  10.1007/JHEP02(2015)157} {\bibfield  {journal} {\bibinfo  {journal} {JHEP}\
  }\textbf {\bibinfo {volume} {1502}},\ \bibinfo {pages} {157} (\bibinfo {year}
  {2015})},\ \Eprint {http://arxiv.org/abs/1412.0018} {arXiv:1412.0018
  [hep-ph]} \BibitemShut {NoStop}%
\bibitem [{\citenamefont {Khachatryan}\ \emph {et~al.}(2014)\citenamefont
  {Khachatryan} \emph {et~al.}}]{Khachatryan:2014rra}%
  \BibitemOpen
  \bibfield  {author} {\bibinfo {author} {\bibfnamefont {V.}~\bibnamefont
  {Khachatryan}} \emph {et~al.} (\bibinfo {collaboration} {CMS
  Collaboration}),\ }\href@noop {} {\  (\bibinfo {year} {2014})},\ \Eprint
  {http://arxiv.org/abs/1408.3583} {arXiv:1408.3583 [hep-ex]} \BibitemShut
  {NoStop}%
\bibitem [{\citenamefont {Batell}\ \emph {et~al.}(2009)\citenamefont {Batell},
  \citenamefont {Pospelov},\ and\ \citenamefont {Ritz}}]{Batell:2009di}%
  \BibitemOpen
  \bibfield  {author} {\bibinfo {author} {\bibfnamefont {B.}~\bibnamefont
  {Batell}}, \bibinfo {author} {\bibfnamefont {M.}~\bibnamefont {Pospelov}}, \
  and\ \bibinfo {author} {\bibfnamefont {A.}~\bibnamefont {Ritz}},\ }\href
  {\doibase 10.1103/PhysRevD.80.095024} {\bibfield  {journal} {\bibinfo
  {journal} {Phys.Rev.}\ }\textbf {\bibinfo {volume} {D80}},\ \bibinfo {pages}
  {095024} (\bibinfo {year} {2009})},\ \Eprint {http://arxiv.org/abs/0906.5614}
  {arXiv:0906.5614 [hep-ph]} \BibitemShut {NoStop}%
\bibitem [{\citenamefont {deNiverville}\ \emph {et~al.}(2011)\citenamefont
  {deNiverville}, \citenamefont {Pospelov},\ and\ \citenamefont
  {Ritz}}]{deNiverville:2011it}%
  \BibitemOpen
  \bibfield  {author} {\bibinfo {author} {\bibfnamefont {P.}~\bibnamefont
  {deNiverville}}, \bibinfo {author} {\bibfnamefont {M.}~\bibnamefont
  {Pospelov}}, \ and\ \bibinfo {author} {\bibfnamefont {A.}~\bibnamefont
  {Ritz}},\ }\href {\doibase 10.1103/PhysRevD.84.075020} {\bibfield  {journal}
  {\bibinfo  {journal} {Phys.Rev.}\ }\textbf {\bibinfo {volume} {D84}},\
  \bibinfo {pages} {075020} (\bibinfo {year} {2011})},\ \Eprint
  {http://arxiv.org/abs/1107.4580} {arXiv:1107.4580 [hep-ph]} \BibitemShut
  {NoStop}%
\bibitem [{\citenamefont {deNiverville}\ \emph {et~al.}(2012)\citenamefont
  {deNiverville}, \citenamefont {McKeen},\ and\ \citenamefont
  {Ritz}}]{deNiverville:2012ij}%
  \BibitemOpen
  \bibfield  {author} {\bibinfo {author} {\bibfnamefont {P.}~\bibnamefont
  {deNiverville}}, \bibinfo {author} {\bibfnamefont {D.}~\bibnamefont
  {McKeen}}, \ and\ \bibinfo {author} {\bibfnamefont {A.}~\bibnamefont
  {Ritz}},\ }\href {\doibase 10.1103/PhysRevD.86.035022} {\bibfield  {journal}
  {\bibinfo  {journal} {Phys.Rev.}\ }\textbf {\bibinfo {volume} {D86}},\
  \bibinfo {pages} {035022} (\bibinfo {year} {2012})},\ \Eprint
  {http://arxiv.org/abs/1205.3499} {arXiv:1205.3499 [hep-ph]} \BibitemShut
  {NoStop}%
\bibitem [{\citenamefont {Dobrescu}\ and\ \citenamefont
  {Frugiuele}(2014)}]{Dobrescu:2014fca}%
  \BibitemOpen
  \bibfield  {author} {\bibinfo {author} {\bibfnamefont {B.~A.}\ \bibnamefont
  {Dobrescu}}\ and\ \bibinfo {author} {\bibfnamefont {C.}~\bibnamefont
  {Frugiuele}},\ }\href {\doibase 10.1103/PhysRevLett.113.061801} {\bibfield
  {journal} {\bibinfo  {journal} {Phys.Rev.Lett.}\ }\textbf {\bibinfo {volume}
  {113}},\ \bibinfo {pages} {061801} (\bibinfo {year} {2014})},\ \Eprint
  {http://arxiv.org/abs/1404.3947} {arXiv:1404.3947 [hep-ph]} \BibitemShut
  {NoStop}%
\bibitem [{\citenamefont {Batell}\ \emph {et~al.}(2014)\citenamefont {Batell},
  \citenamefont {Essig},\ and\ \citenamefont {Surujon}}]{Batell:2014mga}%
  \BibitemOpen
  \bibfield  {author} {\bibinfo {author} {\bibfnamefont {B.}~\bibnamefont
  {Batell}}, \bibinfo {author} {\bibfnamefont {R.}~\bibnamefont {Essig}}, \
  and\ \bibinfo {author} {\bibfnamefont {Z.}~\bibnamefont {Surujon}},\ }\href
  {\doibase 10.1103/PhysRevLett.113.171802} {\bibfield  {journal} {\bibinfo
  {journal} {Phys.Rev.Lett.}\ }\textbf {\bibinfo {volume} {113}},\ \bibinfo
  {pages} {171802} (\bibinfo {year} {2014})},\ \Eprint
  {http://arxiv.org/abs/1406.2698} {arXiv:1406.2698 [hep-ph]} \BibitemShut
  {NoStop}%
\bibitem [{\citenamefont {Markevitch}\ \emph {et~al.}(2004)\citenamefont
  {Markevitch}, \citenamefont {Gonzalez}, \citenamefont {Clowe}, \citenamefont
  {Vikhlinin}, \citenamefont {David} \emph {et~al.}}]{Markevitch:2003at}%
  \BibitemOpen
  \bibfield  {author} {\bibinfo {author} {\bibfnamefont {M.}~\bibnamefont
  {Markevitch}}, \bibinfo {author} {\bibfnamefont {A.}~\bibnamefont
  {Gonzalez}}, \bibinfo {author} {\bibfnamefont {D.}~\bibnamefont {Clowe}},
  \bibinfo {author} {\bibfnamefont {A.}~\bibnamefont {Vikhlinin}}, \bibinfo
  {author} {\bibfnamefont {L.}~\bibnamefont {David}},  \emph {et~al.},\ }\href
  {\doibase 10.1086/383178} {\bibfield  {journal} {\bibinfo  {journal}
  {Astrophys.J.}\ }\textbf {\bibinfo {volume} {606}},\ \bibinfo {pages} {819}
  (\bibinfo {year} {2004})},\ \Eprint {http://arxiv.org/abs/astro-ph/0309303}
  {arXiv:astro-ph/0309303 [astro-ph]} \BibitemShut {NoStop}%
\bibitem [{\citenamefont {Miralda-Escude}(2002)}]{MiraldaEscude:2000qt}%
  \BibitemOpen
  \bibfield  {author} {\bibinfo {author} {\bibfnamefont {J.}~\bibnamefont
  {Miralda-Escude}},\ }\href@noop {} {\bibfield  {journal} {\bibinfo  {journal}
  {Astrophys.J.}\ }\textbf {\bibinfo {volume} {564}},\ \bibinfo {pages} {60}
  (\bibinfo {year} {2002})},\ \Eprint {http://arxiv.org/abs/astro-ph/0002050}
  {arXiv:astro-ph/0002050 [astro-ph]} \BibitemShut {NoStop}%
\bibitem [{\citenamefont {Bionta}\ \emph {et~al.}(1987)\citenamefont {Bionta},
  \citenamefont {Blewitt}, \citenamefont {Bratton}, \citenamefont {Casper},
  \citenamefont {Ciocio} \emph {et~al.}}]{Bionta:1987qt}%
  \BibitemOpen
  \bibfield  {author} {\bibinfo {author} {\bibfnamefont {R.}~\bibnamefont
  {Bionta}}, \bibinfo {author} {\bibfnamefont {G.}~\bibnamefont {Blewitt}},
  \bibinfo {author} {\bibfnamefont {C.}~\bibnamefont {Bratton}}, \bibinfo
  {author} {\bibfnamefont {D.}~\bibnamefont {Casper}}, \bibinfo {author}
  {\bibfnamefont {A.}~\bibnamefont {Ciocio}},  \emph {et~al.},\ }\href
  {\doibase 10.1103/PhysRevLett.58.1494} {\bibfield  {journal} {\bibinfo
  {journal} {Phys.Rev.Lett.}\ }\textbf {\bibinfo {volume} {58}},\ \bibinfo
  {pages} {1494} (\bibinfo {year} {1987})}\BibitemShut {NoStop}%
\bibitem [{\citenamefont {Hirata}\ \emph {et~al.}(1987)\citenamefont {Hirata}
  \emph {et~al.}}]{Hirata:1987hu}%
  \BibitemOpen
  \bibfield  {author} {\bibinfo {author} {\bibfnamefont {K.}~\bibnamefont
  {Hirata}} \emph {et~al.} (\bibinfo {collaboration} {Kamiokande-II}),\ }\href
  {\doibase 10.1103/PhysRevLett.58.1490} {\bibfield  {journal} {\bibinfo
  {journal} {Phys.Rev.Lett.}\ }\textbf {\bibinfo {volume} {58}},\ \bibinfo
  {pages} {1490} (\bibinfo {year} {1987})}\BibitemShut {NoStop}%
\bibitem [{\citenamefont {Izaguirre}\ \emph
  {et~al.}(2014{\natexlab{a}})\citenamefont {Izaguirre}, \citenamefont
  {Krnjaic}, \citenamefont {Schuster},\ and\ \citenamefont
  {Toro}}]{Izaguirre:2014bca}%
  \BibitemOpen
  \bibfield  {author} {\bibinfo {author} {\bibfnamefont {E.}~\bibnamefont
  {Izaguirre}}, \bibinfo {author} {\bibfnamefont {G.}~\bibnamefont {Krnjaic}},
  \bibinfo {author} {\bibfnamefont {P.}~\bibnamefont {Schuster}}, \ and\
  \bibinfo {author} {\bibfnamefont {N.}~\bibnamefont {Toro}},\ }\href@noop {}
  {\  (\bibinfo {year} {2014}{\natexlab{a}})},\ \Eprint
  {http://arxiv.org/abs/1411.1404} {arXiv:1411.1404 [hep-ph]} \BibitemShut
  {NoStop}%
\bibitem [{\citenamefont {Bjorken}\ \emph {et~al.}(1988)\citenamefont
  {Bjorken}, \citenamefont {Ecklund}, \citenamefont {Nelson}, \citenamefont
  {Abashian}, \citenamefont {Church} \emph {et~al.}}]{Bjorken:1988as}%
  \BibitemOpen
  \bibfield  {author} {\bibinfo {author} {\bibfnamefont {J.}~\bibnamefont
  {Bjorken}}, \bibinfo {author} {\bibfnamefont {S.}~\bibnamefont {Ecklund}},
  \bibinfo {author} {\bibfnamefont {W.}~\bibnamefont {Nelson}}, \bibinfo
  {author} {\bibfnamefont {A.}~\bibnamefont {Abashian}}, \bibinfo {author}
  {\bibfnamefont {C.}~\bibnamefont {Church}},  \emph {et~al.},\ }\href
  {\doibase 10.1103/PhysRevD.38.3375} {\bibfield  {journal} {\bibinfo
  {journal} {Phys.Rev.}\ }\textbf {\bibinfo {volume} {D38}},\ \bibinfo {pages}
  {3375} (\bibinfo {year} {1988})}\BibitemShut {NoStop}%
\bibitem [{\citenamefont {Riordan}\ \emph {et~al.}(1987)\citenamefont
  {Riordan}, \citenamefont {Krasny}, \citenamefont {Lang}, \citenamefont
  {De~Barbaro}, \citenamefont {Bodek} \emph {et~al.}}]{Riordan:1987aw}%
  \BibitemOpen
  \bibfield  {author} {\bibinfo {author} {\bibfnamefont {E.}~\bibnamefont
  {Riordan}}, \bibinfo {author} {\bibfnamefont {M.}~\bibnamefont {Krasny}},
  \bibinfo {author} {\bibfnamefont {K.}~\bibnamefont {Lang}}, \bibinfo {author}
  {\bibfnamefont {P.}~\bibnamefont {De~Barbaro}}, \bibinfo {author}
  {\bibfnamefont {A.}~\bibnamefont {Bodek}},  \emph {et~al.},\ }\href {\doibase
  10.1103/PhysRevLett.59.755} {\bibfield  {journal} {\bibinfo  {journal}
  {Phys.Rev.Lett.}\ }\textbf {\bibinfo {volume} {59}},\ \bibinfo {pages} {755}
  (\bibinfo {year} {1987})}\BibitemShut {NoStop}%
\bibitem [{\citenamefont {Bross}\ \emph {et~al.}(1991)\citenamefont {Bross},
  \citenamefont {Crisler}, \citenamefont {Pordes}, \citenamefont {Volk},
  \citenamefont {Errede} \emph {et~al.}}]{Bross:1989mp}%
  \BibitemOpen
  \bibfield  {author} {\bibinfo {author} {\bibfnamefont {A.}~\bibnamefont
  {Bross}}, \bibinfo {author} {\bibfnamefont {M.}~\bibnamefont {Crisler}},
  \bibinfo {author} {\bibfnamefont {S.~H.}\ \bibnamefont {Pordes}}, \bibinfo
  {author} {\bibfnamefont {J.}~\bibnamefont {Volk}}, \bibinfo {author}
  {\bibfnamefont {S.}~\bibnamefont {Errede}},  \emph {et~al.},\ }\href
  {\doibase 10.1103/PhysRevLett.67.2942} {\bibfield  {journal} {\bibinfo
  {journal} {Phys.Rev.Lett.}\ }\textbf {\bibinfo {volume} {67}},\ \bibinfo
  {pages} {2942} (\bibinfo {year} {1991})}\BibitemShut {NoStop}%
\bibitem [{\citenamefont {Davoudiasl}\ \emph {et~al.}(2012)\citenamefont
  {Davoudiasl}, \citenamefont {Lee},\ and\ \citenamefont
  {Marciano}}]{Davoudiasl:2012ig}%
  \BibitemOpen
  \bibfield  {author} {\bibinfo {author} {\bibfnamefont {H.}~\bibnamefont
  {Davoudiasl}}, \bibinfo {author} {\bibfnamefont {H.-S.}\ \bibnamefont {Lee}},
  \ and\ \bibinfo {author} {\bibfnamefont {W.~J.}\ \bibnamefont {Marciano}},\
  }\href {\doibase 10.1103/PhysRevD.86.095009} {\bibfield  {journal} {\bibinfo
  {journal} {Phys.Rev.}\ }\textbf {\bibinfo {volume} {D86}},\ \bibinfo {pages}
  {095009} (\bibinfo {year} {2012})},\ \Eprint {http://arxiv.org/abs/1208.2973}
  {arXiv:1208.2973 [hep-ph]} \BibitemShut {NoStop}%
\bibitem [{\citenamefont {Endo}\ \emph {et~al.}(2012)\citenamefont {Endo},
  \citenamefont {Hamaguchi},\ and\ \citenamefont {Mishima}}]{Endo:2012hp}%
  \BibitemOpen
  \bibfield  {author} {\bibinfo {author} {\bibfnamefont {M.}~\bibnamefont
  {Endo}}, \bibinfo {author} {\bibfnamefont {K.}~\bibnamefont {Hamaguchi}}, \
  and\ \bibinfo {author} {\bibfnamefont {G.}~\bibnamefont {Mishima}},\ }\href
  {\doibase 10.1103/PhysRevD.86.095029} {\bibfield  {journal} {\bibinfo
  {journal} {Phys.Rev.}\ }\textbf {\bibinfo {volume} {D86}},\ \bibinfo {pages}
  {095029} (\bibinfo {year} {2012})},\ \Eprint {http://arxiv.org/abs/1209.2558}
  {arXiv:1209.2558 [hep-ph]} \BibitemShut {NoStop}%
\bibitem [{\citenamefont {Babusci}\ \emph {et~al.}(2013)\citenamefont {Babusci}
  \emph {et~al.}}]{Babusci:2012cr}%
  \BibitemOpen
  \bibfield  {author} {\bibinfo {author} {\bibfnamefont {D.}~\bibnamefont
  {Babusci}} \emph {et~al.} (\bibinfo {collaboration} {KLOE-2 Collaboration}),\
  }\href {\doibase 10.1016/j.physletb.2013.01.067} {\bibfield  {journal}
  {\bibinfo  {journal} {Phys.Lett.}\ }\textbf {\bibinfo {volume} {B720}},\
  \bibinfo {pages} {111} (\bibinfo {year} {2013})},\ \Eprint
  {http://arxiv.org/abs/1210.3927} {arXiv:1210.3927 [hep-ex]} \BibitemShut
  {NoStop}%
\bibitem [{\citenamefont {Adlarson}\ \emph {et~al.}(2013)\citenamefont
  {Adlarson} \emph {et~al.}}]{Adlarson:2013eza}%
  \BibitemOpen
  \bibfield  {author} {\bibinfo {author} {\bibfnamefont {P.}~\bibnamefont
  {Adlarson}} \emph {et~al.} (\bibinfo {collaboration} {WASA-at-COSY
  Collaboration}),\ }\href@noop {} {\  (\bibinfo {year} {2013})},\ \Eprint
  {http://arxiv.org/abs/1304.0671} {arXiv:1304.0671 [hep-ex]} \BibitemShut
  {NoStop}%
\bibitem [{\citenamefont {Abrahamyan}\ \emph {et~al.}(2011)\citenamefont
  {Abrahamyan} \emph {et~al.}}]{Abrahamyan:2011gv}%
  \BibitemOpen
  \bibfield  {author} {\bibinfo {author} {\bibfnamefont {S.}~\bibnamefont
  {Abrahamyan}} \emph {et~al.} (\bibinfo {collaboration} {APEX
  Collaboration}),\ }\href {\doibase 10.1103/PhysRevLett.107.191804} {\bibfield
   {journal} {\bibinfo  {journal} {Phys.Rev.Lett.}\ }\textbf {\bibinfo {volume}
  {107}},\ \bibinfo {pages} {191804} (\bibinfo {year} {2011})},\ \Eprint
  {http://arxiv.org/abs/1108.2750} {arXiv:1108.2750 [hep-ex]} \BibitemShut
  {NoStop}%
\bibitem [{\citenamefont {Merkel}\ \emph {et~al.}(2011)\citenamefont {Merkel}
  \emph {et~al.}}]{Merkel:2011ze}%
  \BibitemOpen
  \bibfield  {author} {\bibinfo {author} {\bibfnamefont {H.}~\bibnamefont
  {Merkel}} \emph {et~al.} (\bibinfo {collaboration} {A1 Collaboration}),\
  }\href {\doibase 10.1103/PhysRevLett.106.251802} {\bibfield  {journal}
  {\bibinfo  {journal} {Phys.Rev.Lett.}\ }\textbf {\bibinfo {volume} {106}},\
  \bibinfo {pages} {251802} (\bibinfo {year} {2011})},\ \Eprint
  {http://arxiv.org/abs/1101.4091} {arXiv:1101.4091 [nucl-ex]} \BibitemShut
  {NoStop}%
\bibitem [{\citenamefont {Bjorken}\ \emph {et~al.}(2009)\citenamefont
  {Bjorken}, \citenamefont {Essig}, \citenamefont {Schuster},\ and\
  \citenamefont {Toro}}]{Bjorken:2009mm}%
  \BibitemOpen
  \bibfield  {author} {\bibinfo {author} {\bibfnamefont {J.~D.}\ \bibnamefont
  {Bjorken}}, \bibinfo {author} {\bibfnamefont {R.}~\bibnamefont {Essig}},
  \bibinfo {author} {\bibfnamefont {P.}~\bibnamefont {Schuster}}, \ and\
  \bibinfo {author} {\bibfnamefont {N.}~\bibnamefont {Toro}},\ }\href {\doibase
  10.1103/PhysRevD.80.075018} {\bibfield  {journal} {\bibinfo  {journal}
  {Phys.Rev.}\ }\textbf {\bibinfo {volume} {D80}},\ \bibinfo {pages} {075018}
  (\bibinfo {year} {2009})},\ \Eprint {http://arxiv.org/abs/0906.0580}
  {arXiv:0906.0580 [hep-ph]} \BibitemShut {NoStop}%
\bibitem [{\citenamefont {Reece}\ and\ \citenamefont
  {Wang}(2009)}]{Reece:2009un}%
  \BibitemOpen
  \bibfield  {author} {\bibinfo {author} {\bibfnamefont {M.}~\bibnamefont
  {Reece}}\ and\ \bibinfo {author} {\bibfnamefont {L.-T.}\ \bibnamefont
  {Wang}},\ }\href {\doibase 10.1088/1126-6708/2009/07/051} {\bibfield
  {journal} {\bibinfo  {journal} {JHEP}\ }\textbf {\bibinfo {volume} {0907}},\
  \bibinfo {pages} {051} (\bibinfo {year} {2009})},\ \Eprint
  {http://arxiv.org/abs/0904.1743} {arXiv:0904.1743 [hep-ph]} \BibitemShut
  {NoStop}%
\bibitem [{\citenamefont {Aubert}\ \emph {et~al.}(2009)\citenamefont {Aubert}
  \emph {et~al.}}]{Aubert:2009cp}%
  \BibitemOpen
  \bibfield  {author} {\bibinfo {author} {\bibfnamefont {B.}~\bibnamefont
  {Aubert}} \emph {et~al.} (\bibinfo {collaboration} {BaBar Collaboration}),\
  }\href {\doibase 10.1103/PhysRevLett.103.081803} {\bibfield  {journal}
  {\bibinfo  {journal} {Phys.Rev.Lett.}\ }\textbf {\bibinfo {volume} {103}},\
  \bibinfo {pages} {081803} (\bibinfo {year} {2009})},\ \Eprint
  {http://arxiv.org/abs/0905.4539} {arXiv:0905.4539 [hep-ex]} \BibitemShut
  {NoStop}%
\bibitem [{\citenamefont {Dent}\ \emph {et~al.}(2012)\citenamefont {Dent},
  \citenamefont {Ferrer},\ and\ \citenamefont {Krauss}}]{Dent:2012mx}%
  \BibitemOpen
  \bibfield  {author} {\bibinfo {author} {\bibfnamefont {J.~B.}\ \bibnamefont
  {Dent}}, \bibinfo {author} {\bibfnamefont {F.}~\bibnamefont {Ferrer}}, \ and\
  \bibinfo {author} {\bibfnamefont {L.~M.}\ \bibnamefont {Krauss}},\
  }\href@noop {} {\  (\bibinfo {year} {2012})},\ \Eprint
  {http://arxiv.org/abs/1201.2683} {arXiv:1201.2683 [astro-ph.CO]} \BibitemShut
  {NoStop}%
\bibitem [{\citenamefont {Dreiner}\ \emph {et~al.}(2014)\citenamefont
  {Dreiner}, \citenamefont {Fortin}, \citenamefont {Hanhart},\ and\
  \citenamefont {Ubaldi}}]{Dreiner:2013mua}%
  \BibitemOpen
  \bibfield  {author} {\bibinfo {author} {\bibfnamefont {H.~K.}\ \bibnamefont
  {Dreiner}}, \bibinfo {author} {\bibfnamefont {J.-F.}\ \bibnamefont {Fortin}},
  \bibinfo {author} {\bibfnamefont {C.}~\bibnamefont {Hanhart}}, \ and\
  \bibinfo {author} {\bibfnamefont {L.}~\bibnamefont {Ubaldi}},\ }\href
  {\doibase 10.1103/PhysRevD.89.105015} {\bibfield  {journal} {\bibinfo
  {journal} {Phys.Rev.}\ }\textbf {\bibinfo {volume} {D89}},\ \bibinfo {pages}
  {105015} (\bibinfo {year} {2014})},\ \Eprint {http://arxiv.org/abs/1310.3826}
  {arXiv:1310.3826 [hep-ph]} \BibitemShut {NoStop}%
\bibitem [{\citenamefont {Essig}\ \emph {et~al.}(2011)\citenamefont {Essig},
  \citenamefont {Schuster}, \citenamefont {Toro},\ and\ \citenamefont
  {Wojtsekhowski}}]{Essig:2010xa}%
  \BibitemOpen
  \bibfield  {author} {\bibinfo {author} {\bibfnamefont {R.}~\bibnamefont
  {Essig}}, \bibinfo {author} {\bibfnamefont {P.}~\bibnamefont {Schuster}},
  \bibinfo {author} {\bibfnamefont {N.}~\bibnamefont {Toro}}, \ and\ \bibinfo
  {author} {\bibfnamefont {B.}~\bibnamefont {Wojtsekhowski}},\ }\href {\doibase
  10.1007/JHEP02(2011)009} {\bibfield  {journal} {\bibinfo  {journal} {JHEP}\
  }\textbf {\bibinfo {volume} {1102}},\ \bibinfo {pages} {009} (\bibinfo {year}
  {2011})},\ \Eprint {http://arxiv.org/abs/1001.2557} {arXiv:1001.2557
  [hep-ph]} \BibitemShut {NoStop}%
\bibitem [{\citenamefont {Battaglieri}\ \emph {et~al.}(2014)\citenamefont
  {Battaglieri}, \citenamefont {Boyarinov}, \citenamefont {Bueltmann},
  \citenamefont {Burkert}, \citenamefont {Celentano} \emph
  {et~al.}}]{Battaglieri:2014hga}%
  \BibitemOpen
  \bibfield  {author} {\bibinfo {author} {\bibfnamefont {M.}~\bibnamefont
  {Battaglieri}}, \bibinfo {author} {\bibfnamefont {S.}~\bibnamefont
  {Boyarinov}}, \bibinfo {author} {\bibfnamefont {S.}~\bibnamefont
  {Bueltmann}}, \bibinfo {author} {\bibfnamefont {V.}~\bibnamefont {Burkert}},
  \bibinfo {author} {\bibfnamefont {A.}~\bibnamefont {Celentano}},  \emph
  {et~al.},\ }\href {\doibase 10.1016/j.nima.2014.12.017} {\bibfield  {journal}
  {\bibinfo  {journal} {Nucl.Instrum.Meth.}\ }\textbf {\bibinfo {volume}
  {A777}},\ \bibinfo {pages} {91} (\bibinfo {year} {2014})},\ \Eprint
  {http://arxiv.org/abs/1406.6115} {arXiv:1406.6115 [physics.ins-det]}
  \BibitemShut {NoStop}%
\bibitem [{\citenamefont {Freytsis}\ \emph {et~al.}(2010)\citenamefont
  {Freytsis}, \citenamefont {Ovanesyan},\ and\ \citenamefont
  {Thaler}}]{Freytsis:2009bh}%
  \BibitemOpen
  \bibfield  {author} {\bibinfo {author} {\bibfnamefont {M.}~\bibnamefont
  {Freytsis}}, \bibinfo {author} {\bibfnamefont {G.}~\bibnamefont {Ovanesyan}},
  \ and\ \bibinfo {author} {\bibfnamefont {J.}~\bibnamefont {Thaler}},\ }\href
  {\doibase 10.1007/JHEP01(2010)111} {\bibfield  {journal} {\bibinfo  {journal}
  {JHEP}\ }\textbf {\bibinfo {volume} {1001}},\ \bibinfo {pages} {111}
  (\bibinfo {year} {2010})},\ \Eprint {http://arxiv.org/abs/0909.2862}
  {arXiv:0909.2862 [hep-ph]} \BibitemShut {NoStop}%
\bibitem [{\citenamefont {Wojtsekhowski}(2009)}]{Wojtsekhowski:2009vz}%
  \BibitemOpen
  \bibfield  {author} {\bibinfo {author} {\bibfnamefont {B.}~\bibnamefont
  {Wojtsekhowski}},\ }\href {\doibase 10.1063/1.3232023} {\bibfield  {journal}
  {\bibinfo  {journal} {AIP Conf.Proc.}\ }\textbf {\bibinfo {volume} {1160}},\
  \bibinfo {pages} {149} (\bibinfo {year} {2009})},\ \Eprint
  {http://arxiv.org/abs/0906.5265} {arXiv:0906.5265 [hep-ex]} \BibitemShut
  {NoStop}%
\bibitem [{\citenamefont {Wojtsekhowski}\ \emph {et~al.}(2012)\citenamefont
  {Wojtsekhowski}, \citenamefont {Nikolenko},\ and\ \citenamefont
  {Rachek}}]{Wojtsekhowski:2012zq}%
  \BibitemOpen
  \bibfield  {author} {\bibinfo {author} {\bibfnamefont {B.}~\bibnamefont
  {Wojtsekhowski}}, \bibinfo {author} {\bibfnamefont {D.}~\bibnamefont
  {Nikolenko}}, \ and\ \bibinfo {author} {\bibfnamefont {I.}~\bibnamefont
  {Rachek}},\ }\href@noop {} {\  (\bibinfo {year} {2012})},\ \Eprint
  {http://arxiv.org/abs/1207.5089} {arXiv:1207.5089 [hep-ex]} \BibitemShut
  {NoStop}%
\bibitem [{\citenamefont {Beranek}\ \emph {et~al.}(2013)\citenamefont
  {Beranek}, \citenamefont {Merkel},\ and\ \citenamefont
  {Vanderhaeghen}}]{Beranek:2013yqa}%
  \BibitemOpen
  \bibfield  {author} {\bibinfo {author} {\bibfnamefont {T.}~\bibnamefont
  {Beranek}}, \bibinfo {author} {\bibfnamefont {H.}~\bibnamefont {Merkel}}, \
  and\ \bibinfo {author} {\bibfnamefont {M.}~\bibnamefont {Vanderhaeghen}},\
  }\href {\doibase 10.1103/PhysRevD.88.015032} {\bibfield  {journal} {\bibinfo
  {journal} {Phys.Rev.}\ }\textbf {\bibinfo {volume} {D88}},\ \bibinfo {pages}
  {015032} (\bibinfo {year} {2013})},\ \Eprint {http://arxiv.org/abs/1303.2540}
  {arXiv:1303.2540 [hep-ph]} \BibitemShut {NoStop}%
\bibitem [{\citenamefont {Chavarria}\ \emph {et~al.}(2014)\citenamefont
  {Chavarria}, \citenamefont {Tiffenberg}, \citenamefont {Aguilar-Arevalo},
  \citenamefont {Amidei}, \citenamefont {Bertou} \emph
  {et~al.}}]{Chavarria:2014ika}%
  \BibitemOpen
  \bibfield  {author} {\bibinfo {author} {\bibfnamefont {A.}~\bibnamefont
  {Chavarria}}, \bibinfo {author} {\bibfnamefont {J.}~\bibnamefont
  {Tiffenberg}}, \bibinfo {author} {\bibfnamefont {A.}~\bibnamefont
  {Aguilar-Arevalo}}, \bibinfo {author} {\bibfnamefont {D.}~\bibnamefont
  {Amidei}}, \bibinfo {author} {\bibfnamefont {X.}~\bibnamefont {Bertou}},
  \emph {et~al.},\ }\href@noop {} {\  (\bibinfo {year} {2014})},\ \Eprint
  {http://arxiv.org/abs/1407.0347} {arXiv:1407.0347 [physics.ins-det]}
  \BibitemShut {NoStop}%
\bibitem [{\citenamefont {Cushman}\ \emph {et~al.}(2013)\citenamefont
  {Cushman}, \citenamefont {Galbiati}, \citenamefont {McKinsey}, \citenamefont
  {Robertson}, \citenamefont {Tait} \emph {et~al.}}]{Cushman:2013zza}%
  \BibitemOpen
  \bibfield  {author} {\bibinfo {author} {\bibfnamefont {P.}~\bibnamefont
  {Cushman}}, \bibinfo {author} {\bibfnamefont {C.}~\bibnamefont {Galbiati}},
  \bibinfo {author} {\bibfnamefont {D.}~\bibnamefont {McKinsey}}, \bibinfo
  {author} {\bibfnamefont {H.}~\bibnamefont {Robertson}}, \bibinfo {author}
  {\bibfnamefont {T.}~\bibnamefont {Tait}},  \emph {et~al.},\ }\href@noop {} {\
   (\bibinfo {year} {2013})},\ \Eprint {http://arxiv.org/abs/1310.8327}
  {arXiv:1310.8327 [hep-ex]} \BibitemShut {NoStop}%
\bibitem [{\citenamefont {Gerbier}\ \emph {et~al.}(2014)\citenamefont
  {Gerbier}, \citenamefont {Giomataris}, \citenamefont {Magnier}, \citenamefont
  {Dastgheibi}, \citenamefont {Gros} \emph {et~al.}}]{Gerbier:2014jwa}%
  \BibitemOpen
  \bibfield  {author} {\bibinfo {author} {\bibfnamefont {G.}~\bibnamefont
  {Gerbier}}, \bibinfo {author} {\bibfnamefont {I.}~\bibnamefont {Giomataris}},
  \bibinfo {author} {\bibfnamefont {P.}~\bibnamefont {Magnier}}, \bibinfo
  {author} {\bibfnamefont {A.}~\bibnamefont {Dastgheibi}}, \bibinfo {author}
  {\bibfnamefont {M.}~\bibnamefont {Gros}},  \emph {et~al.},\ }\href@noop {} {\
   (\bibinfo {year} {2014})},\ \Eprint {http://arxiv.org/abs/1401.7902}
  {arXiv:1401.7902 [astro-ph.IM]} \BibitemShut {NoStop}%
\bibitem [{\citenamefont {Kahn}\ \emph {et~al.}(2014)\citenamefont {Kahn},
  \citenamefont {Krnjaic}, \citenamefont {Thaler},\ and\ \citenamefont
  {Toups}}]{Kahn:2014sra}%
  \BibitemOpen
  \bibfield  {author} {\bibinfo {author} {\bibfnamefont {Y.}~\bibnamefont
  {Kahn}}, \bibinfo {author} {\bibfnamefont {G.}~\bibnamefont {Krnjaic}},
  \bibinfo {author} {\bibfnamefont {J.}~\bibnamefont {Thaler}}, \ and\ \bibinfo
  {author} {\bibfnamefont {M.}~\bibnamefont {Toups}},\ }\href@noop {} {\
  (\bibinfo {year} {2014})},\ \Eprint {http://arxiv.org/abs/1411.1055}
  {arXiv:1411.1055 [hep-ph]} \BibitemShut {NoStop}%
\bibitem [{\citenamefont {Izaguirre}\ \emph
  {et~al.}(2014{\natexlab{b}})\citenamefont {Izaguirre}, \citenamefont
  {Krnjaic}, \citenamefont {Schuster},\ and\ \citenamefont
  {Toro}}]{Izaguirre:2014dua}%
  \BibitemOpen
  \bibfield  {author} {\bibinfo {author} {\bibfnamefont {E.}~\bibnamefont
  {Izaguirre}}, \bibinfo {author} {\bibfnamefont {G.}~\bibnamefont {Krnjaic}},
  \bibinfo {author} {\bibfnamefont {P.}~\bibnamefont {Schuster}}, \ and\
  \bibinfo {author} {\bibfnamefont {N.}~\bibnamefont {Toro}},\ }\href {\doibase
  10.1103/PhysRevD.90.014052} {\bibfield  {journal} {\bibinfo  {journal}
  {Phys.Rev.}\ }\textbf {\bibinfo {volume} {D90}},\ \bibinfo {pages} {014052}
  (\bibinfo {year} {2014}{\natexlab{b}})},\ \Eprint
  {http://arxiv.org/abs/1403.6826} {arXiv:1403.6826 [hep-ph]} \BibitemShut
  {NoStop}%
\bibitem [{\citenamefont {Andreas}\ \emph {et~al.}(2013)\citenamefont
  {Andreas}, \citenamefont {Donskov}, \citenamefont {Crivelli}, \citenamefont
  {Gardikiotis}, \citenamefont {Gninenko} \emph {et~al.}}]{Andreas:2013lya}%
  \BibitemOpen
  \bibfield  {author} {\bibinfo {author} {\bibfnamefont {S.}~\bibnamefont
  {Andreas}}, \bibinfo {author} {\bibfnamefont {S.}~\bibnamefont {Donskov}},
  \bibinfo {author} {\bibfnamefont {P.}~\bibnamefont {Crivelli}}, \bibinfo
  {author} {\bibfnamefont {A.}~\bibnamefont {Gardikiotis}}, \bibinfo {author}
  {\bibfnamefont {S.}~\bibnamefont {Gninenko}},  \emph {et~al.},\ }\href@noop
  {} {\  (\bibinfo {year} {2013})},\ \Eprint {http://arxiv.org/abs/1312.3309}
  {arXiv:1312.3309 [hep-ex]} \BibitemShut {NoStop}%
\end{thebibliography}%
\end{document}